\documentclass[vecphys]{svmult}

\usepackage{makeidx}         
\usepackage{graphicx}        
\usepackage{multicol}        
\usepackage{cite}            
\usepackage[bottom]{footmisc}

\makeindex             

\usepackage{color}
\usepackage{bm}
\newcommand{\be}{\begin{equation}}
\newcommand{\ee}{\end{equation}}
\newcommand{\ba}{\begin{eqnarray}}
\newcommand{\ea}{\end{eqnarray}}
\newcommand{\ff}[1]{{\bm #1}}
\newcommand{\tr}{\mbox{tr}}
\newcommand{\Tr}{\mbox{Tr}}
\newcommand{\refeq}[1]{Eq.\ (\ref{eq:#1})}
\newcommand{\labeq}[1]{\label{eq:#1}}
\newcommand{\reffig}[1]{Fig.\ \ref{fig:#1}}

\newcommand{\bi}{\begin{itemize}}
\newcommand{\ei}{\end{itemize}}

\begin{document}

\title{Self-energy-functional theory}
\author{Michael Potthoff}
\institute{I. Institut f\"ur Theoretische Physik, Universit\"at Hamburg, Jungiusstr. 9, 20355 Hamburg, Germany
\texttt{michael.potthoff@physik.uni-hamburg.de}
}

\maketitle

\begin{abstract} 
Self-energy-functional theory is a formal framework which allows to derive non-perturbative and thermodynamically consistent approximations for lattice models of strongly correlated electrons from a general dynamical variational principle. 
The construction of the self-energy functional and the corresponding variational principle is developed within the path-integral formalism.
Different cluster mean-field approximations, like the variational cluster approximation and cluster extensions of dynamical mean-field theory are derived in this context and their mutual relationship and internal consistency are discussed.
\end{abstract}

\section{Motivation}

The method of Green's functions and diagrammatic perturbation theory \cite{AGD64} represents a powerful approach to study systems of interacting electrons in the thermodynamical limit. 
Several interesting phenomena, like spontaneous magnetic order, correlation-driven metal-insulator transitions or high-temperature superconductivity, however, emerge in systems where electron correlations are strong.
Rather than starting from the non-interacting Fermi gas as the reference point around which the perturbative expansion is developed, a local perspective appears to be more attractive for strongly correlated electron systems,
in particular for prototypical lattice models with local interaction, such as the famous Hubbard model \cite{Hub63,Gut63,Kan63}:
\be
  H = \sum_{ij\sigma} t_{ij} c^\dagger_{i\sigma}c_{j\sigma} + \frac{U}{2} \sum_{i\sigma} n_{i\sigma} n_{i-\sigma} \: .
\ee
The local part of the problem, i.e.\ the Hubbard atom, can be solved easily since its Hilbert space is small.
It is therefore tempting to start from the atomic limit and to treat the rest of the problem, the ``embedding'' of the atom into the lattice, in some approximate way. 
The main idea of the so-called Hubbard-I approximation \cite{Hub63} is to calculate the one-electron Green's function from the Dyson equation where the self-energy is approximated by the self-energy of the atomic system.
This is one of the most simple embedding procedures. 
It already shows that the language of diagrammatic perturbation theory, Green's functions and diagrammatic objects, such as the self-energy, can be very helpful to construct an embedding scheme.

The Hubbard-I approach turns out to be a too crude approximation to describe the above-mentioned collective phenomena. 
One of its advantages, however, is that it offers a perspective for systematic improvement:
Nothing prevents us to start with a more complicated ``atom'' and employ the same trick:
We consider a partition of the underlying lattice with $L$ sites (where $L \to \infty$) into $L/L_c$ disconnected clusters consisting of $L_c$ sites each. 
If $L_c$ is not too large, the self-energy of a single Hubbard cluster is accessible by standard numerical means \cite{Dag94} and can be used as an approximation in the Dyson equation to get the Green's function of the full model. 
This leads to the cluster perturbation theory (CPT) \cite{GV93,SPPL00}. \index{cluster perturbation theory}

CPT can also be motivated by treating the Hubbard interacton $U$ and the {\em inter}-cluster hopping $V$ as a perturbation of the system of disconnected clusters with {\em intra}-cluster hopping $t'$. 
The CPT Green's function is then obtained by summing the diagrams in perturbation theory to all orders in $U$ and $V$ but neglecting vertex corrections which intermix $U$ and $V$ interactions. 

While these two ways of deriving CPT are equivalent, one aspect of the former is interesting: 
Taking the self-energy from some reference model (the cluster) is reminiscent of dynamical mean-field theory (DMFT) \cite{MV89,GKKR96,KV04} where the self-energy of an impurity model approximates the self-energy of the lattice model. 
This provokes the question whether both, the CPT and the DMFT, can be understood in single unifying theoretical framework.
\index{dynamical mean-field theory}

This question is one motivation for the topic of this chapter on self-energy-functional theory (SFT) \cite{Pot03a,Pot03b,PAD03,Pot05}.
Another one is that there are certain deficiencies of the CPT.
While CPT can be seen as a cluster mean-field approach since correlations beyond the cluster extensions are neglected, it not self-consistent, i.e.\ there is no feedback of the resulting Green's function on the cluster to be embedded
(some {\em ad hoc} element of self-consistency is included in the original Hubbard-I approximation).
\index{cluster mean-field approach}
In particular, there is no concept of a Weiss mean field and, therefore, CPT cannot describe different phases of a thermodynamical system nor phase transitions.
Another related point is that CPT provides the Green's function only but no thermodynamical potential.
Different ways to derive e.g.\ the free energy from the Green's function \cite{AGD64,FW71,NO88} give inconsistent results.

To overcome these deficiencies, a self-consistent cluster-embedding scheme has to be set up. 
Ideally, this results from a variational principle for a general thermodynamical potential which is formulated in terms of dynamical quantities as e.g.\ the self-energy or the Green's function.
The variational formulation should ensure the internal consistency of corresponding approximations and should make contact with the DMFT. 
This sets the goals of self-energy-functional theory and also the plan of this chapter. 

\section{Self-energy functional}

\subsection{Hamiltonian, grand potential and self-energy}

We consider a system of electrons in thermodynamical equilibrium at temperature $T$ and chemical potential $\mu$. 
The Hamiltonian of the system $H = H(\ff t, \ff U) = H_0(\ff t) + H_{1}(\ff U)$ consists of a non-interacting part specified by one-particle parameters $\ff t$ and an interaction part with interaction parameters $\ff U$: 
\ba
  H_0(\ff t) &=& \sum_{\alpha\beta} t_{\alpha\beta} \:
  c^\dagger_\alpha c_\beta \; ,
\nonumber \\  
  H_{1}(\ff U) &=& \frac{1}{2} 
  \sum_{\alpha\beta\gamma\delta} U_{\alpha\beta\delta\gamma} \:
  c^\dagger_\alpha c^\dagger_\beta c_\gamma c_\delta \: .
\ea
The index $\alpha$ refers to an arbitrary set of quantum numbers labelling an orthonormal basis of one-particle states $|\alpha\rangle$. 
As is apparent from the form of $H$, the total particle number $N$ is conserved. 

The grand potential of the system is given by $\Omega_{\ff t,\ff U} = - T \ln Z_{\ff t,\ff U}$ where
\be
  Z_{\ff t,\ff U} = \tr \rho_{\ff t,\ff U} \quad \mbox{with} \quad \; \rho_{\ff t,\ff U} = \exp(-(H(\ff t, \ff U) - \mu N)/T) 
\labeq{partf}
\ee
is the partition function. 
The dependence of the partition function (and of other quantities discussed below) on the parameters $\ff t$ and $\ff U$ is made explicit through the subscripts.

The one-particle Green's function $G_{\alpha\beta}(\omega) \equiv \langle\langle c_\alpha ; c_\beta^\dagger \rangle \rangle$ of the system is the main object of interest. 
It will provide the static expectation value of the one-particle density matrix $c^\dagger_\alpha c_\beta$ and the spectrum of one-particle excitations related to a photoemission experiment \cite{Pot01b}.
The Green's function can be defined for complex $\omega$ via its spectral representation:
\be
  G_{\alpha\beta}(\omega) = \int_{-\infty}^\infty dz \: \frac{A_{\alpha\beta}(z)}{\omega -  z} \: ,
\labeq{spectralrep}
\ee
where the spectral density $A_{\alpha\beta}(z) = \int_{-\infty}^\infty dt \: \exp(izt) A_{\alpha\beta}(t)$ is the Fourier tranform of 
\be
  A_{\alpha\beta}(t) = \frac{1}{2\pi} \langle [c_\alpha(t) , c^\dagger_\beta(0)]_+ \rangle \; ,
\ee
which involves the anticommutator of an annihilator and a creator with a Heisenberg time dependence $O(t) = \exp(i(H-\mu N)t) O \exp(- i(H-\mu N)t)$.

Due to the thermal average, $\langle {\cal O} \rangle =  \tr (\rho_{\ff t, \ff U} {\cal O}) / Z_{\ff t,\ff U}$, the Green's function depends on $\ff t$ and $\ff U$ and is denoted by $G_{\ff t, \ff U ,\alpha\beta}(\omega)$.
For the diagram technique employed below, we need the Green's function on the imaginary Matsubara frequencies $\omega = i \omega_n \equiv i(2n+1)\pi T$ with integer $n$ \cite{AGD64}.
In the following the elements $G_{\ff t, \ff U ,\alpha\beta}(i\omega)$ are considered to form a matrix $\ff G_{\ff t, \ff U}$ which is diagonal with respect to $n$.

The ``free'' Green's function $\ff G_{\ff t, 0}$ is obtained for $\ff U = 0$, and its elements are given by:
\be
  G_{\ff t, 0, \alpha\beta}(i\omega_n) = 
  \left(\frac{1}{i\omega_n + \mu - \ff t}\right)_{\alpha\beta} \: .
\ee
Therewith, we can define the self-energy via Dyson's equation
\be
  \ff G_{\ff t, \ff U} = \frac{\ff 1}{\ff G_{\ff t, 0}^{-1} - \ff \Sigma_{\ff t, \ff U}} \; , 
\labeq{dyson}
\ee
i.e.\ $\ff \Sigma_{\ff t, \ff U} = \ff G_{\ff t, 0}^{-1} - \ff G_{\ff t, \ff U}^{-1}$.
The full meaning of this definition becomes clear within the context of diagrammatic perturbation theory \cite{AGD64}.

Here, we like to list some important properties of the self-energy only: 
(i) Via Dyson's equation, it determines the Green's function.
(ii) The self-energy has a spectral representation similar to \refeq{spectralrep}.
(iii) In particular, the coresponding spectral function (matrix) is positive definite, and the poles of
$\ff \Sigma_{\ff t, \ff U}$ are on the real axis \cite{Lut61}.
(iv) $\Sigma_{\alpha\beta}(\omega) = 0$ if $\alpha$ or $\beta$ refer to one-particle orbitals that are non-interacting, i.e.\ if $\alpha$ or $\beta$ do not occur as an entry of the matrix of interaction parameters $\ff U$. 
Those orbitals or sites are called non-interacting. 
This property of the self-energy is clear from its diagrammatic representation.
(v) If $\alpha$ refers to the sites of a Hubbard-type model with local interaction, the self-energy can  generally be assumed to be more local than the Green's function. 
This is corroborated e.g.\ by explicit calculations using weak-coupling perturbation theory \cite{SC90,SC91,PN97c} and by the fact that the self-energy is purely local on infinite-dimensional lattices \cite{MV89,MH89b}.

\subsection{Luttinger-Ward functional}

\index{Luttinger-Ward functional}
We would like to distuinguish between dynamic quantities, like the self-energy, which is frequency-dependent and related to the (one-particle) excitation spectrum, on the one hand, and static quantities, like the grand potential and its derivatives with respect to $\mu$, $T$, etc.\ which are related to the thermodynamics, on the other. 
A link between static and dynamic quantities is needed to set up a variational principle which gives the (dynamic) self-energy by requiring a (static) thermodynamical potential be stationary, 
There are several such relations \cite{AGD64,FW71,NO88}.
The Luttinger-Ward (LW) functional $\widehat{\Phi}_{\ff U}[\ff G]$ provides a special relation with several advantageous properties \cite{LW60}: 
\bi
\item[](i) 
$\widehat{\Phi}_{\ff U}[\ff G]$ is a functional.
Functionals $\widehat{A} = \widehat{A}[\cdots]$ are indicated by a hat and should be distinguished clearly from physical quantities $A$.
\\

\item[](ii)
The domain of the LW functional is given by ``the space of Green's functions''. 
This has to be made more precise later. 
\\

\item[](iii)
If evaluated at the exact (physical) Green's function, $\ff G_{\ff t, \ff U}$, of the system with Hamiltonian $H = H(\ff t, \ff U)$, the LW functional gives a quantity
\be
  \widehat{\Phi}_{\ff U}[\ff G_{\ff t, \ff U}] = \Phi_{\ff t, \ff U}
\labeq{phiphys}
\ee
which is related to the grand potential of the system via:
\be
  \Omega_{\ff t, \ff U}  
  = 
  \Phi_{\ff t, \ff U} 
  +
  \Tr \ln \ff G_{\ff t, \ff U} 
  - 
  \Tr ( \ff \Sigma_{\ff t, \ff U} \ff G_{\ff t, \ff U}) \; .
\labeq{phiom}
\ee
Here the notation $\Tr \ff A \equiv T \sum_n \sum_\alpha e^{i\omega_n0^+} A_{\alpha\alpha}(i\omega_n)$ is used.
$0^+$ is a positive infinitesimal.
\\

\item[](iv)
The functional derivative of the LW functional with respect to its argument is:
\be
  \frac{1}{T} \frac{\delta \widehat{\Phi}_{\ff U}[\ff G]}{\delta \ff G} 
  = \widehat{\ff \Sigma}_{\ff U}[\ff G] \: .
\label{eq:der}
\ee
Clearly, the result of this operation is a functional of the Green's function again. 
This functional is denoted by $\widehat{\ff \Sigma}$ since its evaluation at the physical (exact) Green's function $\ff G_{\ff t, \ff U}$ yields the physical self-energy:
\be
  \widehat{\ff \Sigma}[\ff G_{\ff t, \ff U}] = \ff \Sigma_{\ff t, \ff U}  \: .
\labeq{sigmaskel}
\ee

\item[](v)
The LW functional is ``universal'': 
The functional relation $\widehat{\Phi}_{\ff U}[\cdots]$ is completely determined by the interaction parameters $\ff U$ (and does not depend on $\ff t$).
This is made explicit by the subscript.
Two systems (at the same chemical potential $\mu$ and temperature $T$) with the same interaction $\ff U$ but different one-particle parameters $\ff t$ (on-site energies and hopping integrals) are described by the same Luttinger-Ward functional. 
Using Eq.\ (\ref{eq:der}), this implies that the functional $\widehat{\ff \Sigma}_{\ff U}[\ff G]$ is universal, too.
\\

\item[](vi)
Finally, the LW functional vanishes in the non-interacting limit:
\be
  \widehat{\Phi}_{\ff U}[\ff G] \equiv 0  \quad \mbox{for} \quad \ff U = 0 \; .
\ee
\ei

\subsection{Diagrammatic derivation}

\begin{figure}[t]
  \centerline{\includegraphics[width=0.55\textwidth]{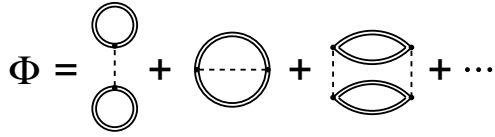}}
\caption{
Original definition of the Luttinger-Ward functional $\widehat{\Phi}_{\ff U}[\ff G]$, see Ref.\ \cite{LW60}.
Double lines: fully interacting propagator $\ff G$. 
Dashed lines: interaction $\ff U$.
}
\label{fig:lw}
\end{figure}

\index{skeleton diagram expansion}
In the original paper by Luttinger and Ward \cite{LW60} it is shown that $\widehat{\Phi}_{\ff U}[\ff G]$ can be constructed order by order in diagrammatic perturbation theory.
The functional is obtained as the limit of the infinite series of closed renormalized skeleton diagrams, i.e.\
closed diagrams without self-energy insertions and with all free propagators replaced by fully interacting ones (see Fig.\ \ref{fig:lw}). 
There is no known case where this skeleton-diagram expansion could be summed up to get a closed form for $\widehat{\Phi}_{\ff U}[\ff G]$.
Therefore, the explicit functional dependence is unknown even for the most simple types of interactions like the Hubbard interaction.

Using the classical diagrammatic definition of the LW functional, the properties (i) -- (vi) listed in the previous section are easily verified:
By construction, $\widehat{\Phi}_{\ff U}[\ff G]$ is a functional of $\ff G$ which is unisersal (properties (i), (ii), (v)).
Any diagram is the series depends on $\ff U$ and on $\ff G$ {\em only}.
Particularly, it is independent of $\ff t$.
Since there is no zeroth-order diagram, $\widehat{\Phi}_{\ff U}[\ff G]$ trivially vanishes for 
$\ff U=0$, this proves (vi).

Diagrammatically, the functional derivative of $\widehat{\Phi}_{\ff U}[\ff G]$ with respect to $\ff G$ corresponds to the removal of a propagator from each of the $\Phi$ diagrams. 
Taking care of topological factors \cite{LW60,AGD64}, one ends up with the skeleton-diagram expansion of the self-energy (iv).
Therefore, Eq.\ (\ref{eq:sigmaskel}) is obtained in the limit of this expansion. 

\refeq{phiom} can be derived by a coupling-constant integration \cite{LW60}.
Alternatively, it can be verified by integrating over $\mu$:
We note that the $\mu$ derivative of the l.h.s and of the r.h.s of \refeq{phiom} are equal for any fixed 
interaction strength $\ff t$, $\ff U$ and $T$. 
Namely,
$(\partial / \partial \mu) (\Phi_{\ff t, \ff U} + \Tr \ln \ff G_{\ff t, \ff U} 
- \Tr \, \ff \Sigma_{\ff t, \ff U} \ff G_{\ff t, \ff U}) =  
\Tr \, \ff G_{\ff t, \ff U}^{-1} (\partial \ff G_{\ff t, \ff U} / \partial \mu) 
- \Tr \, \ff G_{\ff t, \ff U} (\partial \ff \Sigma_{\ff t, \ff U} / 
\partial \mu) = - \Tr \: \ff G_{\ff t, \ff U} = - \langle N \rangle_{\ff t, \ff U} 
= \partial \Omega_{\ff t, \ff U} / \partial \mu$. 
Here, we have used \refeq{partf} in the last step and \refeq{dyson}, \refeq{phiphys} and \refeq{der} before. 
$\langle N \rangle_{\ff t, \ff U}$ is the grand-canonical average of the total particle-number operator.
Integration over $\mu$ then yields \refeq{phiom}.
Note that the equation holds trivially for $\mu \to -\infty$, i.e.\ for 
$\langle N \rangle_{\ff t, \ff U} \to 0$ since $\ff \Sigma_{\ff t, \ff U} = 0$ 
and $\Phi_{\ff t, \ff U} =0$ in this limit.

\subsection{Derivation using the path integral}

\index{fermion path integral}
For the diagrammatic derivation it has to be assumed that the skeleton-diagram series is convergent. 
It is therefore interesting to see how the LW functional can be defined and how its properties can be verified within a path-integral formulation.
This is non-perturbative.
The path-integral construction of the LW functional was first given in Ref.\ \cite{Pot06b}.

Using Grassmann variables \cite{NO88}
$\xi_\alpha(i\omega_n)= T^{1/2} \int_0^{1/T} d\tau \: e^{i\omega_n\tau} \xi_\alpha(\tau)$ and $\xi^\ast_\alpha(i\omega_n)= T^{1/2} \int_0^{1/T} d\tau \: e^{-i\omega_n\tau} \xi^\ast_\alpha(\tau)$, the elements of the Green's function are given by 
$G_{\ff t, \ff U, \alpha\beta}(i\omega_n) = - \langle \xi_\alpha(i\omega_n) \xi^\ast_\beta(i\omega_n) \rangle_{\ff t, \ff U}$. 
The average
\be
  G_{\ff t, \ff U, \alpha\beta}(i\omega_n) 
  =
  \frac{-1}{Z_{\ff t,\ff U}}
  \int D \xi D \xi^\ast 
  \xi_\alpha(i\omega_n) \xi^\ast_\beta(i\omega_n)
  \exp\left( A_{\ff t, \ff U,\xi\xi^\ast} \right)
\labeq{gfunct}
\ee
is defined with the help of the action
$A_{\ff t, \ff U,\xi\xi^\ast} = A_{\ff t,\xi\xi^\ast} + A_{\ff U,\xi\xi^\ast}$
where
\be
   A_{\ff t, \ff U,\xi\xi^\ast} 
   =
   \sum_{n,\alpha\beta} \xi_\alpha^\ast(i\omega_n) 
   ((i\omega_n + \mu)\delta_{\alpha\beta} - t_{\alpha\beta})
   \xi_\beta(i\omega_n) 
   +
   A_{\ff U,\xi\xi^\ast}
\ee
and 
\be
   A_{\ff U,\xi\xi^\ast}
   = 
   - \frac{1}{2} \sum_{\alpha\beta\gamma\delta} U_{\alpha\beta\delta\gamma} \int_0^{1/T} 
   \!\! d\tau \:
   \xi_\alpha^\ast(\tau) 
   \xi_\beta^\ast(\tau)
   \xi_\gamma(\tau) 
   \xi_\delta(\tau) \; .
\ee
This is the standard path-integral representation of the Green's function \cite{NO88}.

The action can be considered as the physical action that is obtained when evaluating the {\em functional}
\be
   \widehat{A}_{\ff U,\xi\xi^\ast}[\ff G_0^{-1}] =
   \sum_{n,\alpha\beta} \xi_\alpha^\ast(i\omega_n) 
   G_{0,\alpha\beta}^{-1}(i\omega_n)
   \xi_\beta(i\omega_n) 
   +
   A_{\ff U,\xi\xi^\ast} \: 
\ee
at the (matrix inverse of the) physical free Green's function, i.e.\
\be
   A_{\ff t, \ff U,\xi\xi^\ast} 
   = 
   \widehat{A}_{\ff U,\xi\xi^\ast}[\ff G_{\ff t,0}^{-1}] \; .
\ee
Using this, we define the functional
\be
  \widehat{\Omega}_{\ff U}[\ff G_0^{-1}] =  - T \ln \widehat{Z}_{\ff U}[\ff G_0^{-1}]
\labeq{omgfree}
\ee
with
\be
  \widehat{Z}_{\ff U}[\ff G_0^{-1}] = \int D \xi D \xi^\ast 
  \exp\left( \widehat{A}_{\ff U,\xi\xi^\ast}[\ff G_0^{-1}] \right) \; .
\ee
The functional dependence of $\widehat{\Omega}_{\ff U}[\ff G_0^{-1}]$ is determined by $\ff U$ only, i.e.\ the functional is universal. 
Obviously, the physical grand potential is obtained when inserting the physical inverse free Green's function
$\ff G_{\ff t, 0}^{-1}$:
\be
  \widehat{\Omega}_{\ff U}[\ff G_{\ff t, 0}^{-1}] = \Omega_{\ff t, \ff U} \: .
\labeq{omex0} 
\ee
The functional derivative of \refeq{omgfree} leads to another universal functional:
\be
\frac{1}{T} \frac{\delta \widehat{\Omega}_{\ff U}[\ff G_0^{-1}]}
{\delta \ff G_0^{-1}} 
= - \: \frac{1}{\widehat{Z}_{\ff U}[\ff G_0^{-1}]} 
\frac{\delta \widehat{Z}_{\ff U}[\ff G_0^{-1}]}
{\delta \ff G_0^{-1}} 
\equiv - \widehat{\cal \ff G}_{\ff U}[\ff G_0^{-1}] \: ,
\labeq{omder}
\ee
with the property
\be
  \widehat{\cal \ff G}_{\ff U}[\ff G_{\ff t, 0}^{-1}] = \ff G_{\ff t, \ff U} \: .
\labeq{calg}
\ee
This is easily seen from \refeq{gfunct}.

The strategy to be pursued is the following: 
$\widehat{\cal \ff G}_{\ff U}[\ff G_{0}^{-1}]$ is a universal ($\ff t$ independent) functional and can be used to construct a universal relation $\ff \Sigma = \widehat{\ff \Sigma}_{\ff U}[\ff G]$ 
between the self-energy and the one-particle Green's function -- independent from the Dyson equation (\ref{eq:dyson}).
Using this and the universal functional $\widehat{\Omega}_{\ff U}[\ff G_{0}^{-1}]$, a universal functional $\widehat{\Phi}_{\ff U}[\ff G]$ can be constructed, the derivative of which essentially yields $\widehat{\ff \Sigma}_{\ff U}[\ff G]$ and that also obeys all other properties of the diagrammatically constucted LW functional.

To start with, consider the equation
\be
  \widehat{\cal \ff G}_{\ff U}[ \ff G^{-1} + \ff \Sigma ] = \ff G \: .
\labeq{rel}  
\ee
This is a relation between the variables $\ff \Sigma$ and $\ff G$ which
for a given $\ff G$, may be solved for $\ff \Sigma$. 
This defines a functional $\widehat{\ff \Sigma}_{\ff U}[\ff G]$, i.e.
\be
  \widehat{\cal \ff G}_{\ff U}[ \ff G^{-1} - \widehat{\ff \Sigma}_{\ff U}[\ff G] ] = \ff G \: .
\labeq{gfunc}
\ee
For a given Green's function $\ff G$, the self-energy $\ff \Sigma = \widehat{\ff \Sigma}_{\ff U}[\ff G]$ is defined to be the solution of \refeq{rel}.
From the Dyson equation (\ref{eq:dyson}) and \refeq{calg} it is obvious that the relation (\ref{eq:rel}) is satisfied for the physical $\ff \Sigma = \ff \Sigma_{\ff t, \ff U}$ and the physical $\ff G=\ff G_{\ff t, \ff U}$ of the system with Hamiltonian $H_{\ff t, \ff U}$:
\be
  \widehat{\ff \Sigma}_{\ff U}[ \ff G_{\ff t, \ff U} ] = \ff \Sigma_{\ff t, \ff U} \: .
\labeq{gex}
\ee
This construction simplifies the original presentation in Ref.\ \cite{Pot06b}. 
The discussion on the existence and the uniqueness of possible solutions of the relation (\ref{eq:rel}) given there applies accordingly to the present case.

With the help of the functionals $\widehat{\Omega}_{\ff U}[\ff G_0^{-1}]$ and $\widehat{\ff \Sigma}_{\ff U}[\ff G]$, the Luttinger-Ward functional is obtained as:
\be
  \widehat{\Phi}_{\ff U}[\ff G] = 
  \widehat{\Omega}_{\ff U}[ 
  \ff G^{-1} + \widehat{\ff \Sigma}_{\ff U}[\ff G]
  ]
  - 
  \Tr \ln \ff G
  +
  \Tr ( \widehat{\ff \Sigma}_{\ff U}[\ff G] \ff G) \: .
\labeq{phidef}
\ee
Let us check property (iv).
Using \refeq{omder} one finds for the derivative of the first term:
\be
  \frac{1}{T} 
  \frac{\delta
  \widehat{\Omega}_{\ff U}[ \ff G^{-1} + \widehat{\ff \Sigma}_{\ff U}[\ff G] ]
  }{\delta G_{\alpha\beta}} 
  =
  - \sum_{\gamma\delta} \widehat{\cal G}_{\ff U,\gamma\delta}[ 
  \ff G^{-1} + \widehat{\ff \Sigma}_{\ff U}[\ff G]
  ] 
  \left(
  \frac{\delta G_{\gamma\delta}^{-1}}{\delta G_{\alpha\beta}} + \frac{\delta \widehat{\Sigma}_{\ff U,\gamma\delta}[\ff G]}{\delta G_{\alpha\beta}} 
  \right)
\ee
and, using \refeq{gfunc},
\be
  \frac{1}{T} 
  \frac{\delta
  \widehat{\Omega}_{\ff U}[ \ff G^{-1} + \widehat{\ff \Sigma}_{\ff U}[\ff G] ]
  }{\delta G_{\alpha\beta}} 
  =
  - \sum_{\gamma\delta} G_{\gamma\delta}
  \left(
  \frac{\delta G_{\gamma\delta}^{-1}}{\delta G_{\alpha\beta}} + \frac{\delta \widehat{\Sigma}_{\ff U,\gamma\delta}[\ff G]}{\delta G_{\alpha\beta}} 
  \right) \: .
\ee
Therewith,
\be
\frac{1}{T} \frac{\delta \widehat{\Phi}_{\ff U}[\ff G]}{\delta G_{\alpha\beta}} 
  =
  G_{\beta\alpha}^{-1}
  -
  \sum_{\gamma\delta } G_{\gamma\delta} 
  \frac{\delta \widehat{\Sigma}_{\ff U,\gamma\delta}[\ff G]}{\delta G_{\alpha\beta}}
  +
  \frac{1}{T} \frac{\delta}{\delta G_{\alpha\beta}} 
  \left( - 
  \Tr \ln \ff G
  +
  \Tr ( \widehat{\ff \Sigma}_{\ff U}[\ff G] \ff G) 
  \right)
\ee
and, finally, 
\be
\frac{1}{T} \frac{\delta \widehat{\Phi}_{\ff U}[\ff G]}{\delta G_{\alpha\beta}(i\omega_n)} 
=
\widehat{\Sigma}_{\ff U, \beta\alpha}(i\omega_n)[\ff G] \: ,
\label{eq:pder}
\ee
where, as a reminder, the frequency dependence has been reintroduced.

The other properties are also verified easily. 
(i) and (ii) are obvious.
(iii) follows from \refeq{omex0}, \refeq{calg} and \refeq{gex} and the Dyson equation (\ref{eq:dyson}).
The universality of the LW functional (v) is ensured by the presented construction. 
It involves universal functionals only.
Finally, (vi) follows from 
$\widehat{\cal \ff G}_{\ff U=0}[\ff G^{-1}] = \ff G$
which implies (via \refeq{gfunc})
$\widehat{\ff \Sigma}_{\ff U=0}[\ff G] = 0$, and with 
$\widehat{\Omega}_{\ff U=0}[\ff G^{-1}] = \Tr \ln \ff G$
we get
$\widehat{\Phi}_{\ff U=0}[\ff G] = 0$.

\subsection{Variational principle}

The functional $\ff \Sigma_{\ff U}[\ff G]$ can be assumed to be invertible {\em locally} provided that the system is not at a critical point for a phase transition (see also Ref.\ \cite{Pot03a}). 
This allows to construct the Legendre transform of the LW functional:
\be
  \widehat{F}_{\ff U}[\ff \Sigma] = \widehat{\Phi}_{\ff U}[\widehat{\ff G}_{\ff U}[\ff \Sigma]] 
  - 
  \Tr (\ff \Sigma \widehat{\ff G}_{\ff U}[\ff \Sigma]) \: .
\ee
Here, $\widehat{\ff G}_{\ff U}[\widehat{\ff \Sigma}_{\ff U}[\ff G]] = \ff G$. 
With \refeq{pder} we immediately find
\be
\frac{1}{T} \frac{\delta \widehat{F}_{\ff U}[\ff \Sigma]}{\delta \ff \Sigma} 
=
- \widehat{\ff G}_{\ff U}[\ff \Sigma] \: .
\label{eq:fder}
\ee
We now define the self-energy functional:
\be
  \widehat{\Omega}_{\ff t, \ff U}[\ff \Sigma] = 
  \Tr \ln \frac{1}{\ff G_{\ff t,0}^{-1} - \ff \Sigma}
  + \widehat{F}_{\ff U}[\ff \Sigma] \: .
\labeq{sef}
\ee
Its functional derivative is easily calculated:
\be
  \frac{1}{T} \frac{\delta \widehat{\Omega}_{\ff t, \ff U}[\ff \Sigma]}
  {\delta \ff \Sigma} = 
  \frac{1}{\ff G_{\ff t,0}^{-1} - \ff \Sigma} - 
  \widehat{\ff G}_{\ff U}[\ff \Sigma] \: .
\ee
The equation 
\be
  \widehat{\ff G}_{\ff U}[\ff \Sigma] = \frac{1}{\ff G_{\ff t,0}^{-1} - \ff \Sigma}
\label{eq:sig}
\ee
is a (highly non-linear) conditional equation for the self-energy of the system $H = H_0(\ff t) + H_{1}(\ff U)$.
Eqs.\ (\ref{eq:dyson}) and (\ref{eq:gex}) show that it is satisfied by the physical self-energy $\ff \Sigma = \ff \Sigma_{\ff t, \ff U}$.
Note that the l.h.s of (\ref{eq:sig}) is independent of $\ff t$ but depends on $\ff U$ (universality of $\widehat{\ff G}_{\ff U}[\ff \Sigma]$), while the r.h.s  is independent of $\ff U$ but depends on $\ff t$ via $\ff G_{\ff t,0}^{-1}$.
The obvious problem of finding a solution of \refeq{sig} is that there is no closed form for the functional $\widehat{\ff G}_{\ff U}[\ff \Sigma]$.
Solving Eq.\ (\ref{eq:sig}) is equivalent to a search for the stationary point of the grand potential as a functional of the self-energy:
\be
  \frac{\delta \widehat{\Omega}_{\ff t, \ff U}[\ff \Sigma]}
  {\delta \ff \Sigma} = 0 \; .
\label{eq:var}
\ee
This is the starting point for self-energy-functional theory.

\subsection{Approximation schemes}
\label{sec:types}

Up to this point we have discussed exact relations only. 
It is clear, however, that it is generally impossible to evaluate the self-energy functional \refeq{sef} for a given $\ff \Sigma$ and that one has to resort to approximations.
Three different types of approximation strategies may be distinguished:

In a {\em type-I approximation} one derives the Euler equation
$\delta \widehat{\Omega}_{\ff t, \ff U}[\ff \Sigma] / \delta \ff \Sigma = 0$ first and then chooses (a physically motivated) simplification of the equation afterwards to render the determination of $\ff \Sigma_{\ff t, \ff U}$ possible.
This is most general but also questionable a priori, as normally the approximated Euler equation no longer derives from some approximate functional.
This may result in thermodynamical inconsistencies.

A {\em type-II approximation} modifies the form of the functional dependence,
$\widehat{\Omega}_{\ff t, \ff U}[\cdots] \to \widehat{\Omega}^{(1)}_{\ff t, \ff U}[\cdots]$, to get a simpler one that allows for a solution of the resulting Euler equation $\delta \widehat{\Omega}^{(1)}_{\ff t, \ff U}[\ff \Sigma] / \delta \ff \Sigma = 0$.
This type is more particular and yields a thermodynamical potential consistent with $\ff \Sigma_{\ff t, \ff U}$.
Generally, however, it is not easy to find a sensible approximation of a functional form.

Finally, in a {\em type-III approximation} one restricts the domain of the functional which must then be defined precisely.
This type is most specific and, from a conceptual point of view, should be preferred as compared to type-I or type-II approximations as the exact functional form is retained.
In addition to conceptual clarity and thermodynamical consistency, type-III approximations are truely systematic since improvements can be obtained by an according extension of the domain.

Examples for the different cases can be found e.g.\ in Ref.\ \cite{Pot05}.
The classification of approximation schemes is hierarchical:
Any type-III approximation can always be understood as a type-II one, and any type-II approximations as type-I, but not vice versa.
This does not mean, however, that type-III approximations are superior as compared to type-II and type-I ones.
They are conceptually more appealing but do not necessarily provide ``better'' results.
One reason to consider self-energy functionals instead of functionals of the Green's function (see Refs.\ \cite{CK00,CK01}, for example), is to derive the DMFT as a type-III approximation.

\section{Variational cluster approach} 

The central idea of self-energy-functional theory is to make use of the universality of (the Legendre transform of) the Luttinger-Ward functional to construct type-III approximations.
Consider the self-energy functional \refeq{sef}. 
Its first part consists of a simple explicit functional of $\ff \Sigma$ while its second part, i.e.\ $\widehat{F}_{\ff U}[\ff \Sigma]$, is unknown but depends on $\ff U$ only.

\subsection{Reference system}

Due to this universality of $\widehat{F}_{\ff U}[\ff \Sigma]$, one has
\be
  \widehat{\Omega}_{\ff t', \ff U}[\ff \Sigma] = 
  \Tr \ln \frac{1}{\ff G_{\ff t',0}^{-1} - \ff \Sigma}
  + \widehat{F}_{\ff U}[\ff \Sigma] \: 
\labeq{sfp}
\ee
for the self-energy functional of a so-called ``reference system''. 
As compared to the original system of interest, the reference system is given by a Hamiltonian $H' = H_{0} (\ff t') + H_{1} (\ff U)$ with the same interaction part $\ff U$ but modified one-particle parameters $\ff t'$.
The reference system has different microscopic parameters but is assumed to be in the 
same macroscopic state, i.e.\ at the same temperature $T$ and the same chemical 
potential $\mu$.
By a proper choice of its one-particle part, the problem posed by the reference system $H'$ can be much simpler than the original problem posed by $H$. 
We assume that the self-energy of the reference system $\ff \Sigma_{\ff t',\ff U}$ can be computed exactly, e.g.\ by some numerical technique.

Combining Eqs.\ (\ref{eq:sef}) and (\ref{eq:sfp}), one can eliminate the unknown functional $\widehat{F}_{\ff U}[\ff \Sigma]$:
\be
  \widehat{\Omega}_{\ff t, \ff U}[\ff \Sigma] 
  = 
  \widehat{\Omega}_{\ff t', \ff U}[\ff \Sigma] 
  + 
  \Tr \ln \frac{1}{\ff G_{\ff t,0}^{-1} - \ff \Sigma}
  - 
  \Tr \ln \frac{1}{\ff G_{\ff t',0}^{-1} - \ff \Sigma} \: .
  \labeq{sfp1}
\ee
It appears that this amounts to a shift of the problem only as the self-energy {\em functional} of the reference system again contains the full complexity of the problem.
In fact, except for the trivial case $\ff U=0$, the functional dependence of $\widehat{\Omega}_{\ff t', \ff U}[\ff \Sigma]$ is unknown -- even if the reference system is assumed to be solvable, i.e.\ if the self-energy $\ff \Sigma_{\ff t',\ff U}$, the Green's function $\ff G_{\ff t',\ff U}$ and the grand potential $\Omega_{\ff t', \ff U}$ of the reference system are available.

However, inserting the self-energy of the reference system $\ff \Sigma_{\ff t',\ff U}$ into the self-energy functional of the original one, and using $\widehat{\Omega}_{\ff t', \ff U}[\ff \Sigma_{\ff t',\ff U}] = \Omega_{\ff t',\ff U}$ and the Dyson equation of the reference system,
we find:
\be
  \widehat{\Omega}_{\ff t, \ff U}[\ff \Sigma_{\ff t',\ff U}] 
  = 
  {\Omega}_{\ff t', \ff U}
  + 
  \Tr \ln \frac{1}{\ff G_{\ff t,0}^{-1} - \ff \Sigma_{\ff t',\ff U}}
  - 
  \Tr \ln \ff G_{\ff t',\ff U} \:  .
\labeq{ocalc}	
\ee
This is a remarkable result.
It shows that an {\em exact} evaluation of the self-energy functional of a non-trivial original system is possible, at least for certain self-energies.
This requires to solve a reference system with the same interaction part.

\subsection{Domain of the self-energy functional}

\begin{figure}[t]
  \centerline{\includegraphics[width=0.5\textwidth]{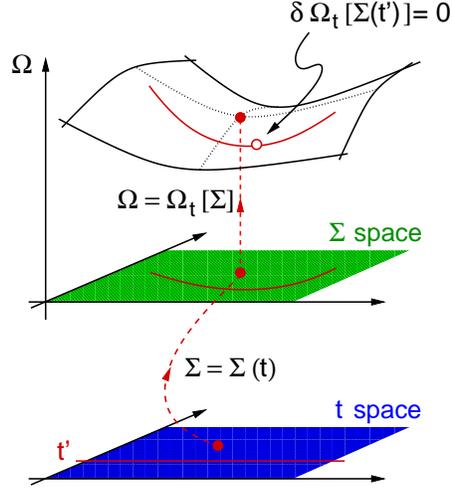}}
\caption{
Schematic illustration for the construction of consistent approximations within the SFT. 
The grand potential is considered as a functional of the self-energy which is parametrized by the one-particle parameters $\ff t$ of the Hamiltonian ($\ff U$ is fixed).
At the physical self-energy, $\Omega$ is stationary (filled red circles).
The functional dependence of $\Omega$ on $\ff \Sigma$ is not accessible in the entire space of self-energies ($\Sigma$ space) but on a restricted subspace of ``trial'' self-energies parametrized by a subset of one-particle parameters $\ff t'$ (solid red lines) corresponding to a ``reference system'', i.e.\ a manifold of systems with the same interaction part but a one-particle part given by $\ff t'$ which renders the solution possible.
The grand potential can be evaluated exactly on the submanifold of reference systems. 
A stationary point on this submanifold represents the approximate self-energy and the corresponding approximate grand potential (open circle).
}
\label{fig:sft}
\end{figure}

\refeq{ocalc} provides an explict expression of the self-energy functional $\widehat{\Omega}_{\ff t, \ff U}[\ff \Sigma]$.
This is suitable to discuss the domain of the functional precisely. 
Take $\ff U$ to be fixed.
We define the space of $\ff t$-representable self-energies as 
\be
  {\cal S}_{\ff U} 
  = 
  \{
  \ff \Sigma \: | \: \exists \ff t: \; \ff \Sigma = \ff \Sigma_{\ff t,\ff U}
  \}
  \: .
\ee
This definition of the domain is very convenient since it ensures the correct analytical and causal properties of the variable $\ff \Sigma$.

We can now formulate the result of the preceeding section in the following way. 
Consider a set of reference systems with $\ff U$ fixed but different one-particle parameters $\ff t'$, i.e.\ a space of one-particle parameters ${\cal T'}$. 
Assume that the reference system with $H' = H_{\ff t',\ff U}$ can be solved exactly for any $\ff t' \in {\cal T'}$.
Then, the self-energy functional $\widehat{\Omega}_{\ff t, \ff U}[\ff \Sigma]$ can be evaluated exactly on the subspace 
\be
  {\cal S}'_{\ff U} 
  = 
  \{
  \ff \Sigma \: | \: \exists \ff t' \in {\cal T'}: \; \ff \Sigma = \ff \Sigma_{\ff t',\ff U}
  \}
  \subset 
  {\cal S}_{\ff U} 
  \: .
\ee
This fact can be used to construct type-III approximations, see \reffig{sft}.

\subsection{Construction of cluster approximations} 

\begin{figure}[b]
\centering
\includegraphics[width=0.4\columnwidth]{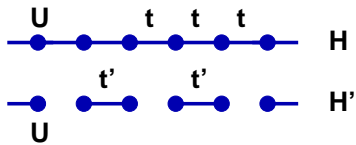}
\caption{
Variational cluster approximation (VCA) for the Hubbard model. 
Top: representation of the original one-dimensional Hubbard model $H$ with nearest-neighbor hopping $t$ and Hubbard-interaction $U$.
Bottom: reference system $H'$ consisting of decoupled clusters of $L_c=2$ sites each with intra-cluster hopping $t'$ as a variational parameter.
}
\label{fig:refsys}
\end{figure}

A certain approximation is defined by a choice of the reference system or actually by a manifold of reference systems specified by a manifold of one-particle parameters ${\cal T}'$.
As an example consider \reffig{refsys}. 
The original system is given by the one-dimensional Hubbard model with nearest-neighbor hopping $t$ and Hubbard interaction $U$.
A possible reference system is given by switching off the hopping between clusters consisting of $L_c=2$ sites each.
The hopping within the cluster $t'$ is arbitrary, this defines the space ${\cal T}'$.
The self-energies in ${\cal S}'$, the corresponding Green's functions and grand potentials of the reference system can obviously be calculated easily since the degrees of freedom are decoupled spatially. 
Inserting these quantities in \refeq{ocalc} yields the SFT grand potential as a function of $\ff t'$:
\be
  \Omega(\ff t') \equiv \widehat{\Omega}_{\ff t, \ff U}[\ff \Sigma_{\ff t',\ff U}] \: .
\labeq{sftgp}
\ee
This is no longer a functional but an ordinary function of the variational parameters $\ff t' \in {\cal T}'$.
The final task then consists in finding a stationary point $\ff t'_{\rm opt}$ of this function:
\be
  \frac{\partial \Omega(\ff t')}{\partial \ff t'} = 0 \qquad \mbox{for} \; \ff t' = \ff t'_{\rm opt} \: .
\labeq{stat}
\ee

In the example considered this is a function of a single variable $t'$ (we assume $t'$ to be the same for all clusters).
Note that not only the reference system (in the example the isolated cluster) defines the final result but also the lattice structure and the one-particle parameters of the original system.
These enter $\Omega(\ff t')$ via the free Green's function $G_{\ff t,0}$ of the original system.
In the first term on the r.h.s of \refeq{ocalc} we just recognize the CPT Green's function
$1/(\ff G_{\ff t,0}^{-1} - \ff \Sigma_{\ff t',\ff U})$.
The approximation generated by a reference system of disconnected clusters is called variational cluster approximation (VCA). 

\begin{figure}[t]
\centering
\includegraphics[width=0.5\columnwidth]{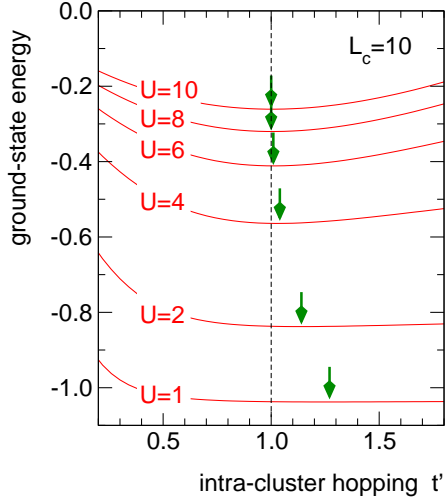}
\caption{
(taken from Ref.\ \cite{BHP08}).
SFT ground-state energy per site, i.e.\ $(\Omega(t')+\mu \langle N\rangle)/L$, as a function of the intra-cluster nearest-neighbor hopping $t'$ for $L_c=10$ and different $U$ ($\mu=U/2$) at zero temperature. 
Arrows indicate the respective optimal $t'$.
The energy scale is fixed by $t=1$.
}
\label{fig:omegat}
\end{figure}

An example for the results of a numerical calculation is given in Fig.\ \ref{fig:omegat}, see also Ref.\ \cite{BHP08}.
The calculation has been performed for the one-dimensional particle-hole symmetric Hubbard model at half-filling and zero temperature.
The figure shows the numerical results for the optimal nearest-neighbor intra-cluster hopping $t'$ as obtained from the VCA for a reference system with disconnected clusters consisting of $L_c=10$ sites each.
The hopping $t'$ is assumed to be the same for all pairs of nearest neighbors.
In principle, one could vary all one-particle parameters that do not lead to a coupling of the clusters to get the optimal result.
In most cases, however, is it necessary to restrict oneself to a small number of physically motivated variational parameters to avoid complications arising from a search for a stationary point in a high-dimensional parameter space.
For the example discussed here, the parameter space ${\cal T}'$ is one-dimensional only.
This is the most simple choice but more elaborate approximations can be generated easily.
The flexibility to construct approximations of different quality and complexity must be seen as one of the advantages of the variational cluster approximation and of the SFT in general.

As can be seen from the figure, a non-trivial result, namely $t'_{\rm opt} \ne t=1$, is found for the optimal value of $t'_{\rm opt}$. 
We also notice that $t_{\rm opt} > t$. 
The physical interpretation is that switching off the {\em inter}-cluster hopping, which generates the approximate self-energy, can partially be compensated for by enhancing the {\em intra}-cluster hopping.
The effect is the more pronounced the smaller is the cluster size $L_c$.
Furthermore, it is reasonable that in case of a stronger interaction and thus more localized electrons, switching off the inter-cluster hopping is less significant. 
This can be seen in \reffig{omegat}: 
The largest optimal hopping $t'_{\rm opt}$ is obtained for the smallest $U$.

On the other hand, even a ``strong'' approximation for the self-energy (measured as a strong deviation of $t'_{\rm opt}$ from $t$) becomes irrelevant in the weak-coupling limit because the self-energy must vanish for $U=0$.
Generally, we note that the VCA becomes exact in the limit $\ff U = 0$: In \refeq{sfp1} the first and the third terms on r.h.s cancel each other and we are left with 
\be
  \widehat{\Omega}_{\ff t, \ff U=0}[\ff \Sigma] 
  = 
  \Tr \ln \frac{1}{\ff G_{\ff t,0}^{-1} - \ff \Sigma} \; .
  \labeq{sfp10}
\ee
Since the trial self-energy has to be taken from a reference system with the same interaction part, i.e.\ $\ff U=0$ and thus $\ff \Sigma=0$, the limit becomes trivial.
For weak but finite $\ff U$, the SFT grand potential becomes flatter and flatter until for $\ff U=0$ the $\ff t'$ dependence is completely irrelevant.

The VCA is also exact in the atomic limit or, more general and again trivial, in the case that there is no restriction on the trial self-energies: ${\cal S}' = {\cal S}$.
In this case, $\ff t'_{\rm opt} = \ff t$ solves the problem, i.e.\ the second and the third term on the r.h.s of \refeq{ocalc} cancel each other and $\widehat{\Omega}_{\ff t, \ff U}[\ff \Sigma_{\ff t',\ff U}] = {\Omega}_{\ff t', \ff U}$ for $\ff t' = \ff t$.

Cluster-perturbation theory (CPT) can be understood as being identical with the VCA provided that the SFT expression for the grand potential is used and that no parameter optimization at all is performed. 
As can be seen from \reffig{omegat}, there is a gain in binding energy due to the optimization of $t'$, i.e.\ $\Omega(t'_{\rm opt}) < \Omega(t)$.
This means that the VCA improves on the CPT result.

\begin{figure}[t]
\centering
\includegraphics[width=0.6\columnwidth]{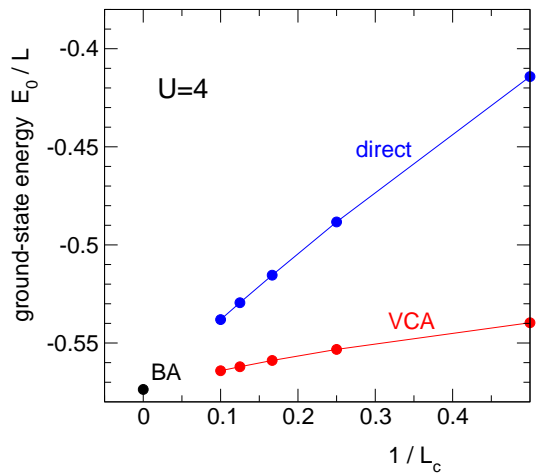}
\caption{
(taken from Ref.\ \cite{BHP08}).
Optimal VCA ground-state energy per site for $U=4$ for different cluster sizes $L_c$ as a function of $1/L_c$ compared to the exact (BA) result and the direct cluster approach.
}
\label{fig:u48}
\end{figure}

\reffig{u48} shows the ground-state energy (per site), i.e.\ the SFT grand potential at zero temperature constantly shifted by $\mu N$ at the stationary point, as a function of the inverse cluster size $1/L_c$.
The dependence turns out to be quite regular and allows to recover the exact Bethe-Ansatz result (BA) \cite{LW68} by extrapolation to $1/L_c = 0$. 
The VCA result represents a considerable improvement as compared to the ``direct'' cluster approach where $E_0$ is simply approximated by the ground-state energy of an isolated Hubbard chain (with open boundary conditions).
Convergence to the exact result is clearly faster within the VCA. 
Note that the direct cluster approach, opposed to the VCA, is not exact for $U=0$.

In the example discussed so far a single variational parameter was taken into account only.
More parameters can be useful for different reasons. 
For example, the optimal self-energy provided by the VCA as a real-space cluster technique artificially breaks the translational symmetry of the original lattice problem. 
Finite-size effects are expected to be the most pronounced at the cluster boundary.
This suggests to consider all intra-cluster hopping parameters as independent variational parameters or at least the hopping at the edges of the chain.

\begin{figure}[t]
\centering
\includegraphics[width=0.45\columnwidth]{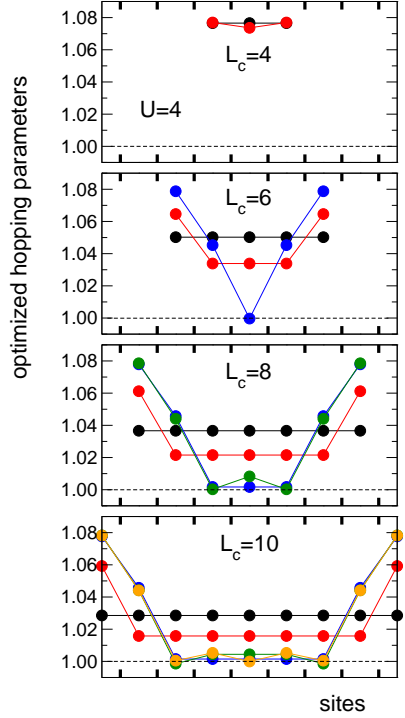}
\caption{
(taken from Ref.\ \cite{BHP08}).
Optimized hopping parameters for the reference systems shown in \reffig{refsys} but for larger clusters with $L_c$ sites each as indicated.
VCA results for $U=4$, $t=1$ at half-filling and temperature $T=0$.
{\em Black}: hopping assumed to be uniform.
{\em Red}: two hopping parameters varied independently,
the hopping at the two cluster edges and the ``bulk'' hopping.
{\em Blue}: hopping at the edges, next to the edges and bulk hopping varied.
{\em Green}: four hopping parameters varied.
{\em Orange}: five hopping parameters varied.
}
\label{fig:hopp}
\end{figure}

The result is shown in \reffig{hopp}.
We find that the optimal hopping varies between different nearest neighbors within a range of less than 10\%. 
At the chain edges the optimal hopping is enhanced to compensate the loss of itinerancy due to the switched-off inter-cluster hopping within the VCA.
With increasing distance to the edges, the hopping quickly decreases.
Quite generally, the third hopping parameter is already close to the physical hopping $t$. 
Looking at the $L_c=10$ results where all (five) different hopping parameters have been varied independently (orange circles), one can see the hopping to slightly oscillate around the bulk value reminiscent of surface Friedel oscillations. 

The optimal SFT grand potential is found to be lower for the inhomogeneous cases as compared to the homogeneous (black) one.
Generally, the more variational parameters are taken into account the higher is the decrease
of the SFT grand potential at optimal parameters.
However, the binding-energy gain due to inhomogeneous hopping parameters is much smaller compared to the gain obtained with a larger cluster. 

Considering an additional hopping parameter $t_{\rm pbc}$ linking the two chain edges as a variational parameter, i.e.\ clusters with periodic boundary conditions always gives a minimal SFT grand potential at $t_{\rm pbc}=0$
(instead of a stationary point at $t_{\rm pbc} = 1$). 
This implies that open boundary conditions are preferred (see also Ref.\ \cite{PAD03}).

\section{Consistency of approximations}

\subsection{Analytical structure of the Green's function}

Constructing approximations within the framework of a dynamical variational principle means that, besides an approximate thermodynamical potential, approximate expressions for the self-energy and the one-particle Greens function are obtained.
This raises the question whether their correct analytical structure is respected in an approximation.
For approximations obtained from self-energy-functional theory this is easily shown to be the case in fact.

The physical self-energy $\Sigma_{\alpha\beta}(\omega)$ and the physical Green's function $G_{\alpha\beta}(\omega)$ are analytical functions in the entire complex $\omega$ plane except for the real axis and have a spectral representation (see \refeq{spectralrep}) with non-negative diagonal elements of the spectral function. 

This trivially holds for the SFT self-energy $\Sigma_{\ff t',\ff U, \alpha\beta}(\omega)$ since by construction $\Sigma_{\ff t',\ff U, \alpha\beta}(\omega)$ is the {\em exact} self-energy of a reference system.
The SFT Green's function is obtained from the SFT self-energy and the free Green's function of the original model via Dyson's equation: 
\be
\ff G = \frac{1}{\ff G_{\ff t,0}^{-1} - \ff \Sigma_{\ff t',\ff U}} \: .
\ee
It is easy to see that it is analytical in the complex plane except for the real axis.
To verify that is has a spectral representation with non-negative spectral function, we can equivalently consider the corresponding retarded quantity
$\ff G_{{\rm ret}} (\omega) = \ff G(\omega + i 0^+)$ for real frequencies $\omega$
and verify that ${\bm G}_{{\rm ret}} = {\bm G}_{\rm R} - i {\bm G}_{\rm I}$ with 
${\bm G}_{\rm R}$, ${\bm G}_{\rm I}$ Hermitian and ${\bm G}_{\rm I}$ non-negative:

We can assume that 
${\bm G}_{0,{\rm ret}} = {\bm G}_{0, \rm R} - i {\bm G}_{0, \rm I}$ with 
${\bm G}_{0, \rm R}$, ${\bm G}_{0, \rm I}$ Hermitian and 
${\bm G}_{0, \rm I}$ non-negative.
Since for Hermitian matrices ${\bm A}$, ${\bm B}$ with ${\bm B}$ non-negative, one has
$1/({\bm A} \pm i {\bm B}) = {\bm X} \mp i {\bm Y}$
with ${\bm X}$, ${\bm Y}$ Hermitian and ${\bm Y}$ non-negative (see Ref.\ \cite{Pot03b}), 
we find
${\bm G}_{0,{\rm ret}}^{-1} = {\bm P}_{\rm R} + i {\bm P}_{\rm I}$
with ${\bm P}_{\rm R}$, ${\bm P}_{\rm I}$ Hermitian and ${\bm P}_{\rm I}$ 
non-negative.
Furthermore, we have 
${\bm \Sigma}_{\rm ret} = {\bm \Sigma}_{\rm R} - i {\bm \Sigma}_{\rm I}$ 
with ${\bm \Sigma}_{\rm R}$, ${\bm \Sigma}_{\rm I}$ Hermitian and 
${\bm \Sigma}_{\rm I}$ non-negative. 
Therefore,
\begin{equation}
  {\bm G}_{\rm ret} 
  = \frac{1}{{\bm P}_{\rm R} + i {\bm P}_{\rm I} 
  - {\bm \Sigma}_{\rm R} + i {\bm \Sigma}_{\rm I}}
  = \frac{1}{{\bm Q}_{\rm R} + i {\bm Q}_{\rm I}}
\end{equation}
with ${\bm Q}_{\rm R}$ Hermitian and ${\bm Q}_{\rm I}$ Hermitian 
and non-negative.

Note that the proof given here is simpler than corresponding proofs for cluster extensions of the DMFT \cite{HMJK00,KSPB01} because the SFT does not involve a ``self-consistency condition'' which is the main object of concern for possible causality violations.

\subsection{Thermodynamical consistency} 

\index{thermodynamical consistency}
An advantegeous feature of the VCA and of other approximations within the SFT framework is their internal thermodynamical consistency. 
This is due to the fact that all quantities of interest are derived from an approximate but explicit expression for a thermodynamical potential. 
In principle the expectation value of any observable should be calculated by via
\be
  \langle A \rangle = \frac{\partial \Omega}{\partial \lambda_A} \; ,
\ee
where $\Omega \equiv \Omega(\ff t')$ is the SFT grand potential (see \refeq{sftgp}) at $\ff t' = \ff t'_{\rm opt}$ and $\lambda_A$ is a parameter in the Hamiltonian of the original system which couples linearly to $A$, i.e.\ $H_{\ff t,\ff U} = H_0 + \lambda_A A$.
This ensures, for example, that the Maxwell relations
\be
  \frac{\partial \langle A \rangle}{\partial \lambda_B} 
  = 
  \frac{\partial \langle B \rangle}{\partial \lambda_A} 
\ee
are respected. 

Furthermore, thermodynamical consistency means that expectation values of arbitrary one-particle operators $A= \sum_{\alpha\beta} A_{\alpha\beta} c^\dagger_\alpha c_\beta$ can consistently either be calculated by a corresponding partial derivative of the grand potential on the one hand, or by integration of the one-particle spectral function on the other. 
As an example we consider the total particle number $N=\sum_\alpha c_\alpha^\dagger c_\alpha$. 
{\em A priori} it not guarateed that in an approximate theory the expressions
\begin{equation}
  \langle N \rangle = - \frac{\partial \Omega}{\partial \mu} \: ,
\label{eq:n1}
\end{equation}  
and
\begin{equation}
  \langle N \rangle = \sum_{\alpha} \int_{-\infty}^\infty f(z) A_{\alpha\alpha}(z) dz \: 
\label{eq:n2}
\end{equation}  
with $f(z) = 1/ (\exp(z/T) + 1)$ and the spectral function $A_{\alpha\beta}(z)$ will give the same result.

To prove thermodynamic consistency, we start from Eq.\ (\ref{eq:n1}).
According to \refeq{ocalc}, there is a twofold $\mu$ dependence of $\Omega \equiv \Omega_{\ff t, \ff U} [\ff \Sigma_{\ff t'_{\rm opt},\ff U}]$:
(i) the {\em explicit} $\mu$ dependence due to the chemical potential in the free Green's function of the original model, $\ff G_{\ff t,0}^{-1} = \omega + \mu - \ff t$,
and (ii) an {\em implicit} $\mu$ dependence due to the $\mu$ dependence of the self-energy $\ff \Sigma_{\ff t'_{\rm opt},\ff U}$, the Green's function $\ff G_{\ff t'_{\rm opt},\ff U}$ and the grand potential $\Omega_{\ff t'_{\rm opt},\ff U}$ of the reference system:
\begin{equation}
  \langle N \rangle 
  = 
  - \frac{\partial \Omega}{\partial \mu_{\rm ex.}} 
  - \frac{\partial \Omega}{\partial \mu_{\rm im.}} \: .
\labeq{exim}
\end{equation}
Note that the implicit $\mu$ dependence is due to the chemical potential of the reference system which, by construction, is in the same macroscopic state as the original system {\em as well as} due to the $\mu$ dependence of the stationary point $\ff t'_{\rm opt}$ itself.
The latter can be ignored since
\begin{equation}
  \frac{\partial \Omega}{\partial \ff t'} 
  \cdot
  \frac{\partial \ff t'}{\partial \mu} = 0
\end{equation}
for $\ff t' = \ff t'_{\rm opt}$ because of stationarity condition \refeq{stat}.
 
We assume that an overall shift of the one-particle energies $\varepsilon' \equiv t'_{\alpha\alpha}$ is included in the set ${\cal T}'$ of variational parameters.
Apart from the sign this is essentially the ``chemical potential'' in the reference system but should be formally distinguished from $\mu$ since the latter has a macroscopic thermodynamical meaning and is the same as the chemical potential of the original system which should not be seen as a variational parameter.

The self-energy, the Green's function and the grand potential of the reference system are defined as grand-canonical averages.
Hence, their $\mu$ dependence due to the grand-canonical Hamiltonian ${\cal H'} = H' - \mu N$ is (apart from the sign) the same as their dependence on $\varepsilon'$:
Consequently, we have:
\begin{equation}
  \frac{\partial \Omega}{\partial \mu_{\rm im.}} 
  =
  - \frac{\partial \Omega}{\partial \varepsilon'}
  = 0
\end{equation}
due to the stationarity condition again.

We are then left with the explicit $\mu$ dependence only:
\be
  \frac{\partial \Omega}{\partial \mu_{\rm ex.}} 
  =
  \frac{\partial}{\partial \mu_{\rm ex.}}
  \Tr \ln \frac{1}{ \ff G_{\ff t,0}^{-1} - \ff \Sigma_{\ff t'_{\rm opt},\ff U}}
  =
  - \Tr \frac{1}{ \ff G_{\ff t,0}^{-1} - \ff \Sigma_{\ff t'_{\rm opt},\ff U}} \; .
\ee
Converting the sum over the Matsubara frequencies implicit in the trace $\Tr$ into a contour integral in the complex $\omega$ plane and using Cauchy's theorem, we can proceed to an integration over real frequencies.
Inserting into Eq.\ (\ref{eq:exim}), this yields:
\begin{equation}
  \langle N \rangle 
   = 
  - \frac{1}{\pi} \mbox{Im}
  \int_{-\infty}^\infty f(\omega) \: \tr
  \frac{1}{\ff G_{0,\ff t}^{-1}
  - \ff \Sigma_{\ff t',\ff U}} \Bigg|_{\omega + i 0^+} d\omega 
\label{eq:nn}  
\end{equation}
for $\ff t' = \ff t'_{\rm opt}$ which is just the average particle number given by (\ref{eq:n2}). 
This completes the proof. 

\subsection{Symmetry breaking}

\index{spontaneous symmetry breaking}
The above discussion has shown that besides intra-cluster hopping parameters it can also be important to treat one-particle energies in the reference cluster as variational parameters. 
In particular, one may consider variational parameters which lead to a lower symmetry of the Hamiltonian. 

As an example consider the Hubbard model on a bipartite lattice as the system of interest and disconnected clusters of size $L_c$ as a reference system. 
The reference-system Hamiltonian shall include, e.g., an additional a staggered magnetic-field term:
\be
  H'_{\rm fict.} = B' \sum_{i\sigma} z_i (n_{i\uparrow} - n_{i\downarrow}) \; ,
\ee
where $z_i = +1$ for sites on sublattice 1, and $z_i = -1$ for sublattice 2. 
The additional term $H'_{\rm fict.}$ leads to a valid reference system as there is no change of the interaction part.
We include the field strength $B'$ in the set of variational parameters, $B' \in {\cal T}'$.

$B'$ is the strength of a ficticious field or, in the language of mean-field theory, the strength of the internal magnetic field or the Weiss field.
This has to be distinguished clearly from an external {\em physical} field applied to the system with field strength $B$:
\be
  H_{\rm phys.} = B \sum_{i\sigma} z_i (n_{i\uparrow} - n_{i\downarrow}) \; 
\ee
This term adds to the Hamiltonian of the original system.

We expect $B'_{\rm opt}=0$ in case of the paramagnetic state and $B=0$ (and this is easily verified numerically). 
Consider the $B'$ and $B$ dependence of the SFT grand potential $\Omega(B',B) = \Omega_B[\ff \Sigma_{B'}]$. 
Here we have suppressed the dependencies on other variational parameters $\ff t'$ and on $\ff t,\ff U$.
Due to the stationarity condition, $\partial \Omega(B',B) / \partial B' = 0$, the optimal Weiss field $B'$ can be considered as a function of $B$, i.e.\ $B'_{\rm opt}=B'(B)$. 
Therefore, we also have:
\be
\frac{d}{dB} \frac{\partial \Omega(B'(B),B)}{\partial B'} = 0 \: .
\labeq{cccond}
\ee
This yields:
\be
\frac{\partial^2 \Omega(B'(B),B)}{\partial {B'}^2}
\frac{d B'(B)}{dB} 
+
\frac{\partial^2 \Omega(B'(B),B)}{\partial B \partial B'}
= 0 \: .
\labeq{cccond1}
\ee
Solving for $dB'/dB$ we find:
\be
\frac{dB'}{dB} 
= -
\left[\frac{\partial^2 \Omega}{\partial {B'}^2}\right]^{-1}
\frac{\partial^2 \Omega}{\partial B \partial B'}
\: .
\labeq{cccond2}
\ee
This clearly shows that $B'=B'_{\rm opt}$ has to be interpreted carefully. 
$B'$ can be much stronger than $B$ if the curvature $\partial^2 \Omega/\partial {B'}^2$ of the SFT functional at the stationary point is small, i.e.\ if the functional is rather flat as it is the case in the limit $U\to 0$, for example.

Form the SFT appoximation for the staggered magnetization, 
\be
  m = \sum_{i\sigma} z_i \langle (n_{i\uparrow} - n_{i\downarrow}) \rangle
     \approx \frac{d}{dB}\Omega(B'(B),B) 
     = \frac{\partial \Omega(B'(B),B) }{\partial B} \; ,
\ee
where the stationarity condition has been used once more, 
we can calculate the susceptibility,
\be
\chi = \frac{dm}{dB} 
= 
\frac{\partial^2 \Omega(B'(B),B)}{\partial B' \partial B} 
\frac{d B'(B)}{dB} 
+
\frac{\partial^2 \Omega(B'(B),B)}{\partial B^2}  \: .
\ee
Using \refeq{cccond2},
\be
\chi
= 
\frac{\partial^2 \Omega}{\partial B^2} 
-
\left( \frac{\partial^2 \Omega}{\partial {B'}^2} \right)^{-1}
\:
\left( \frac{\partial^2 \Omega}{\partial B' \partial B} \right)^2 
\: .
\labeq{chi}
\ee
We see that there are two contributions. 
The first term is due to the explicit $B$ dependence in the SFT grand potential while the second is due to the implicit $B$ dependence via the $B$ dependence of the stationary point.
\refeq{chi} also demonstrates (see Ref.\ \cite{Ede09}) that for the calculation of the paramagnetic susceptibility $\chi$ one may first consider spin-independent variational parameters only to find a stationary point.
This strongly reduces the computational effort \cite{BP10}.
Once a stationary point is found, partial derivatives according to \refeq{chi} have to calculated with spin-dependent parameters $\ff \lambda$ in a single final step.

\begin{figure}[t]
\centering
\includegraphics[width=0.6\columnwidth]{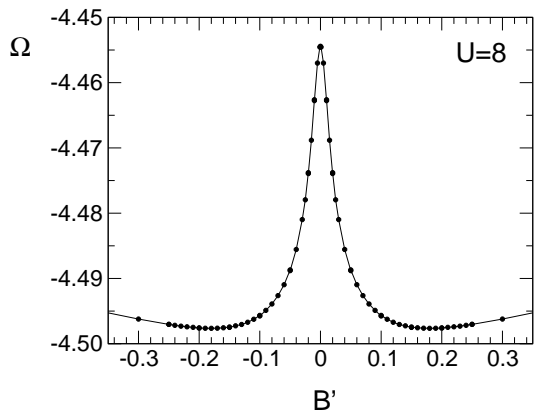}
\caption{
(taken from Ref.\ \cite{DAH+04}).
SFT grand potential as a function of the stength of a ficticious staggered magnetic field $B'$.
VCA calculation using disconnected clusters consisting of $L_c=10$ sites each for the two-dimensional Hubbard model on the square lattice at half-filling, zero temperature, $U=8$ and nearest-neighbor hoppint $t=1$.
}
\label{fig:u8}
\end{figure}

{\em Spontaneous} symmetry breaking is obtained at $B=0$ if there is a stationary point with $B'_{\rm opt} \ne 0$. 
\reffig{u8} gives an example for the particle-hole symmetric Hubbard model on the square lattice at half-filling and zero temperature. 
As a reference system a cluster with $L_c=10$ sites is considered, and the ficticious staggered magnetic field is taken as the only variational parameter.
There is a stationary point at $B'=0$ which corresponds to the paramagnetic phase. 
At $B'=0$ the usual cluster-perturbation theory is recovered.
The two equivalent stationary points at finite $B'$ correspond to a phase with spontaneous antiferromagnetic order -- as expected for the Hubbard model in this parameter regime. 
The antiferromagnetic ground state is stable as compared to the paramagnetic phase.
Its order parameter $m$ is the conjugate variable to the ficticious field. 
Since the latter is a variational parameter, $m$ can either be calculated by integration of the spin-dependent spectral density or as the derivative of the SFT grand potential with respect to the physical field strength $B$ with the same result. 
More details are given in Ref.\ \cite{DAH+04}.

The possibility to study spontaneous symmetry breaking using the VCA with suitably chosen Weiss fields as variational parameters has been exploited frequently in the past. 
Besides antiferromagnetism \cite{DAH+04,NSST08,HKSO08,YO09}, spiral phases \cite{SS08}, ferromagnetism \cite{BP10}, $d$-wave superconductivity \cite{SLMT05,AA05,AAPH06a,AAPH06b,SS06,AADH09,SS09}, charge order \cite{AEvdLP+04,ASE05} and orbtial order \cite{LA09} have been investigated. 
The fact that an explicit expression for a thermodyanmical potential is available allows to study discontinuous transitions and phase separation as well.

\subsection{Non-perturbative conserving approximations}

\index{non-perturbative conserving approximations}
Continuous symmetries of a Hamiltonian imply the existence of conserved quantities:
The conservation of total energy, momentum, angular momentum, spin and particle number is enforced by a not explicitly time-dependent Hamiltonian which is spatially homogeneous and isotropic and invariant under global SU(2) and U(1) gauge transformations.
Approximations may artificially break symmetries and thus lead to unphysical violations of conservations laws.
Baym and Kadanoff \cite{BK61,Bay62} have analyzed under which circumstances an approximation respect the mentioned macroscopic conservation laws.
Within diagrammatic perturbation theory it could be shown that approximations that derive from an explicit but approximate expression for the LW functional $\Phi$ (''$\Phi$-derivable approximations'') are ``conserving''.
Examples for conserving approximations are the Hartree-Fock or the fluctuation-exchange approximation \cite{BK61,BSW89}.

The SFT provides a framework to construct $\Phi$-derivable approximations for correlated lattice models which are {\em non-perturbative}, i.e.\ do not employ truncations of the skeleton-diagram expansion.
Like in weak-coupling conserving approximations, approximations within the SFT are derived from the LW functional, or its Legendre transform $F_{\ff U}[\ff \Sigma]$. 
These are $\Phi$-derivable since any type-III approximation can also be seen as a type-II one, see Section \ref{sec:types}. 

For fermionic lattice models, conservation of energy, particle number and spin have to be considered.
Besides the static thermodynamics, the SFT concentrates on the {\em one-particle} excitations. 
For the approximate one-particle Green's function, however, it is actually simple to prove directly that the above conservation laws are respected.
A short discussion is given in Ref.\ \cite{OBP07}.

\index{Luttinger's sum rule}
At zero temperature $T=0$ there is another non-trivial theorem which is satisfied by any $\Phi$-derivable approximation, namely Luttinger's sum rule \cite{LW60,Lut60}.
This states that at zero temperature the volume in reciprocal space that is enclosed by the Fermi surface is equal to the average particle number.
The original proof of the sum rule by Luttinger and Ward \cite{LW60} is based on the skeleton-diagram expansion of $\Phi$ in the exact theory and is straightforwardly transferred to the case of a $\Phi$-derivable approximation. 
This also implies that other Fermi-liquid properties, such as the linear trend of the specific heat at low $T$ and Fermi-liquid expressions for the $T=0$ charge and the spin susceptibility are respected by a $\Phi$-derivable approximation.

For approximations constructed within the SFT, a different proof has to be found.
One can start with \refeq{ocalc} and perform the zero-temperature limit for an original system (and thus for a reference system) of finite size $L$. 
The different terms in the SFT grand potential then consist of finite sums.
The calculation proceeds by taking the $\mu$-derivative, for $T=0$, on both sides of \refeq{ocalc}. 
This yields the following result (see Ref.\ \cite{OBP07} for details):
\be
  \langle N \rangle = \langle N \rangle' + 2 \sum_{\ff k} \Theta(G_{\ff k}(0))
  - 2 \sum_{\ff k} \Theta(G'_{\ff k}(0)) \: .
\labeq{res}
\ee
Here $\langle N \rangle$ ($\langle N \rangle'$) is the ground-state expectation value of the total particle number $N$ in the original (reference) system, and $G_{\ff k}(0)$ ($G'_{\ff k}(0)$) are the diagonal elements of the one-electron Green's function $\ff G$ at $\omega=0$.
As Luttinger's sum rule reads
\be
  \langle N \rangle = 2 \sum_{\ff k} \Theta(G'_{\ff k}(0)) \: ,
\ee
this implies that, within an approximation constructed within the SFT, the sum rule is satisfied if and only if it is satisfied for the reference system, i.e.\ if $\langle N \rangle'=2 \sum_{\ff k} \Theta(G'_{\ff k}(0))$.
This demonstrates that the theorem is ``propagated'' to the original system irrespective of the approximation that is constructed within the SFT.
This propagation also works in the opposite direction. 
Namely, a possible violation of the exact sum rule for the reference system would 
imply a violation of the sum rule, expressed in terms of approximate quantities, 
for the original system. 

There are no problems to take the thermodynamic limit $L\to \infty$ (if desired) on both sides of \refeq{res}.
The $\ff k$ sums turn into integrals over the unit cell of the reciprocal lattice.
For a $D$-dimensional lattice the $D-1$-dimensional manifold of $\ff k$ points with 
$G_{\ff k}(0)=\infty$ or $G_{\ff k}(0)=0$ form Fermi or Luttinger surfaces, respectively.
Translational symmetry of the original as well as the reference system may be assumed but is not necessary.
In the absence of translational symmetry, however, one has to re-interprete the wave vector $\ff k$ as an index which refers to the elements of the diagonalized Green's function matrix $\ff G$.
The exact sum rule generalizes accordingly but can no longer be expressed in terms of a Fermi surface since there is no reciprocal space. 
It is also valid for the case of a translationally symmetric original Hamiltonian where, due to the choice of a reference system with reduced translational symmetries, such as employed in the VCA, the symmetries of the (approximate) Green's function of the original system are (artificially) reduced.
Since with \refeq{res} the proof of the sum rule is actually shifted to the proof of the sum rule for the reference system only, we are faced with the interesting problem of the validity of the sum rule for a finite cluster.
For small Hubbard clusters with non-degenerate ground state this has been checked numerically with the surprising result that violations of the sum rule appear in certain parameter regimes close to half-filling, see Ref.\ \cite{OBP07}.
This leaves us with the question where the proof of the sum rule fails if applied to a system of finite size.
This is an open problem that has been stated and discussed in Refs.\ \cite{OBP07,KP07} and that is probably related to the breakdown of the sum rule for Mott insulators \cite{Ros06}.

\section{Bath degrees of freedom}

\subsection{Motivation and dynamical impurity approximation}

Within self-energy-functional theory, an approximation is specified by the choice of the reference system. 
The reference system must share the same interaction part with the original model and should be amenable to a (numerically) exact solution.
These requirements restrict the number of conceivable approximations.
So far we have considered a decoupling of degrees of freedom by partitioning a Hubbard-type lattice model into finite Hubbard clusters of $L_c$ sites each, which results in the variational cluster approximation (VCA).

Another element in constructing approximation is to {\em add} degrees of freedom. 
Since the interaction part has to be kept unchanged, the only possibility to do that consists in adding new uncorrelated sites (or ``orbitals''), i.e.\ sites where the Hubbard-$U$ vanishes.
These are called ``bath sites''. 
The coupling of bath sites to the correlated sites with finite $U$ in the reference system via a one-particle term in the Hamiltonian is called ``hybridization''.

\begin{figure}[t]
\includegraphics[width=85mm]{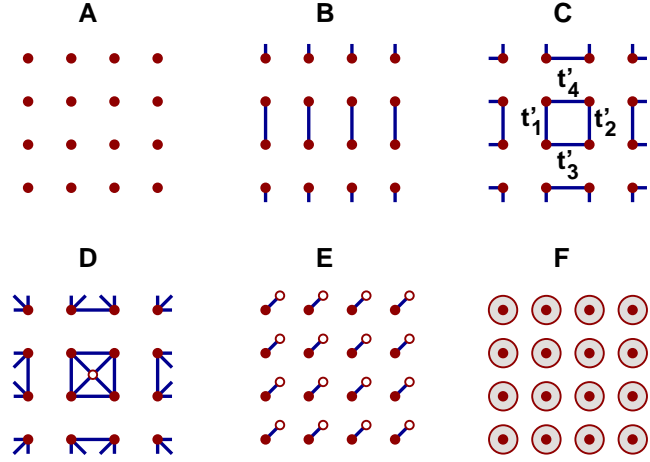}
\centering
\caption{
Different possible reference systems with the same interaction as the single-band Hubbard model on a square lattice.
Filled circles: correlated sites with $U$ as in the Hubbard model.
Open circles: uncorrelated ``bath'' sites with $U=0$.
Lines: nearest-neighbor hopping.
Big circles: continuous bath consisting of $L_{\rm b} = \infty$ bath sites.
Reference systems $H'_{\rm A}$, $H'_{\rm B}$, $H'_{\rm C}$ generate variational cluster approximations (VCA),
$H'_{\rm E}$ yields dynamical impurity approximation (DIA), $H'_{\rm F}$ the DMFT, and
$H'_{\rm D}$ an intermediate approximation (VCA with one additional bath site per cluster).
}
\label{fig:ref}
\end{figure}

\reffig{ref} shows different possibilities.
Reference system A yields a trial self-energy which is local $\Sigma_{ij\sigma}(\omega) = \delta_{ij} \Sigma(\omega)$ and has the same pole structure as the self-energy of the atomic limit of the Hubbard model. 
This results in a variant of the Hubbard-I approximation \cite{Hub63}. 
Reference systems B and C generate variational cluster approximations.
In reference system D an additional bath site is added to the finite cluster.
Reference system E generates a local self-energy again but, as compared to A, allows to treat more variational parameters, namely the on-site energies of the correlated and of the bath site and the hybridization between them. 
We call the resulting approximation a ``dynamical impurity approximation'' (DIA) with $L_b=1$. 

\index{dynamical impurity approximation}
The DIA is a mean-field approximation since the self-energy is local which indicates that non-local two-particle correlations, e.g.\ spin-spin correlations, do not feed back to the one-particle Green function. 
It is, however, quite different from static mean-field (Hartree-Fock) theory since even on the $L_b=1$ level it includes retardation effects that result from processes $\propto V^2$ where the electron hops from the correlated to the bath site and back. 
This improves, as compared to the Hubbard-I approximation, the frequency dependence of the self-energy, i.e.\ the description of the temporal quantum fluctuations. 
The two-site ($L_b=1$) DIA is the most simple approximation which is non-perturbative, conserving, thermodynamically consistent and which respects the Luttinger sum rule (see Ref.\ \cite{OBP07}).
Besides the ``atomic'' physics that leads to the formation of the Hubbard bands, it also includes in the most simple form the possibility to form a local singlet, i.e.\ to screen the local magnetic moment on the correlated site by coupling to the local moment at the bath site.
The correct Kondo scale is missed, of course.
Since the two-site DIA is computationally extremely cheap, is has been employed frequently in the past, in particular to study the physics of the Mott metal-insulator transition \cite{Pot03a,Pot03b,KMOH04,Poz04,IKSK05a,IKSK05b,EKPV07,OBP07}.

\subsection{Relation to dynamical mean-field theory}

Starting from E and adding more and more bath sites to improve the description of temporal fluctuations, one ends up with reference system F where a continuum of bath sites ($L_b=\infty$) is attached to each of the disconnected correlated sites. 
This generates the ``optimal'' DIA.

To characterize this approximation, we consider the SFT grand potential given by \refeq{ocalc} and analyze the stationarity condition \refeq{stat}:
Calculating the derivative with respect to $\ff t'$, we get the general SFT Euler equation:
\be
   T \sum_{n} \sum_{\alpha\beta}
   \left( 
   \frac{1}{{\bf G}_{\ff t,0}^{-1}(i\omega_n) - {\bf \Sigma}_{\ff t',\ff U}(i\omega_n)} 
   -  {\bf G}_{\ff t',\ff U}(i\omega_n) \right)_{\beta \alpha} 
   \frac{\partial \Sigma_{\ff t',\ff U, \alpha\beta}(i\omega_n)}
        {\partial {{\bf t}'}}
   = 0  \: .
\nonumber \\   
\labeq{euler}
\ee
For the physical self-energy $\ff \Sigma_{\ff t,\ff U}$ of the original system $H_{\ff t,\ff U}$, the equation was fulfilled since the bracket would be zero.
Vice versa, since ${\ff G}_{\ff t',\ff U} = \widehat{\ff G}_{\ff U}[{\bf \Sigma}_{\ff t',\ff U}]$, the physical self-energy of $H_{\ff t,\ff U}$ is determined by the condition that the bracket be zero. 
Hence, one can consider the SFT Euler equation to be obtained from the {\em exact} conditional equation for the ``vector'' ${\ff \Sigma}$ in the self-energy space ${\cal S}_{\ff U}$ through {\em projection} onto the hypersurface of 
${\bf t}'$ representable trial self-energies ${\cal S}'_{\ff U}$ by taking the scalar product with vectors 
${\partial \Sigma_{\ff t',\ff U, \alpha\beta}(i\omega_n)}/{\partial {{\bf t}'}}$ tangential to the hypersurface.

Consider now the Hubbard model in particular and the trial self-energies generated by reference system F. 
Actually, F is a set of disconnected single-impurity Anderson models (SIAM).
Assuming translational symmetry, these impurity models are identical replicas.
The self-energy of the SIAM is non-zero on the correlated (``impurity'') site only.
Hence, the trial self-energies are local and site-independent, i.e.\ $\Sigma_{ij\sigma}(i\omega_n) = \delta_{ij} \Sigma(i\omega_n)$, and thus \refeq{euler} reads:
\be
   T \sum_{n} \sum_{i\sigma}
   \left( 
   \frac{1}{{\bf G}_{0}^{-1}(i\omega_n) - {\bf \Sigma}(i\omega_n)} - {\bf G}'(i\omega_n) 
   \right)_{ii\sigma} 
   \frac{\partial \Sigma_{ii\sigma}(i\omega_n)}
   {\partial {{\bf t}'}}
   = 0  \: ,
\labeq{euler1}
\ee
where the notation has been somewhat simplified.
The Euler equation would be solved if one-particle parameters of the SIAM and therewith an impurity self-energy can be found that, when inserted into the Dyson equation of the Hubbard model, yields a Green's function, the {\em local} element of which is equal to the impurity Green's function $\ff G'$. 
Namely, the bracket in \refeq{euler1}, i.e.\ the local (diagonal) elements of the bracket in \refeq{euler}, vanishes. Since the ``projector'' ${\partial \Sigma_{ii\sigma}(i\omega_n)}/{\partial {{\bf t}'}}$ is local, this is sufficient.

This way of solving the Euler equation, however, is just the prescription to obtain the self-energy within dynamical mean-field theory (DMFT) \cite{GKKR96}, and setting the local elements of the bracket to zero is just the self-consistency equation of DMFT.
We therefore see that reference system F generates the DMFT.
It is remarkable that with the VCA and the DMFT quite different approximation can be obtained in one and the same theoretical framework.

Note that for any finite $L_b$, as e.g.\ with reference system E, it is impossible to satisfy the DMFT self-consistency equation exactly since the impurity Green's function $\ff G'$ has a finite number of poles while the lattice Green's function $({1}/{{\bf G}_{0}^{-1}(i\omega_n) - {\bf \Sigma}(i\omega_n)})_{ii\sigma}$ has branch cuts.
Nevertheless, with the ``help'' of the projector, it is easily possible to find a stationary point $\ff t_{\rm opt}$ of the self-energy functional and to satisfy \refeq{euler1}.

Conceptually, this is rather different from the DMFT exact-diagonalization (DMFT-ED) approach \cite{CK94} which also solves a SIAM with a finite $L_b$ but which approximates the DMFT self-consistency condition. 
This means that within DMFT-ED an additional {\em ad hoc} prescription to necessary which, opposed to the DIA, will violate thermodynamical consistency.
However, an algorithmic implementation via a self-consistency cycle to solve the Euler equation is simpler within the DMFT-ED as compared to the DIA \cite{Poz04,EKPV07}. 
It has therefore been suggested to guide improved approximations to the self-consistency condition within DMFT-ED by the DIA \cite{Sen10}.
The convergence of results obtained by the DIA to those of full DMFT with increasing number of bath sites $L_b$ is usually fast.
With respect to the Mott transition in the single-band Hubbard model, $L_b =3$ is sufficient to get almost quantitative agreement with DMFT-QMC results, for example \cite{Poz04}.

\index{real-space DMFT}
Real-space DMFT for inhomogeneous systems \cite{PN95,HCR08,STT+08} with a local but site-dependent self-energy $\Sigma_{ij\sigma}(i\omega_n) = \delta_{ij} \Sigma_i(i\omega_n)$ is obtained from reference system F if the one-particle parameters of the SIAM's at different sites are allowed to be different.
This enlarges the space of trial self-energies.
In this case we get one local self-consistency equation for each site.
The different impurity models can be solved independently from each other in each step of the self-consistency cycle while the coupling between the different sites is provided by the Dyson equation of the Hubbard model.

\subsection{Cluster mean-field approximations}

\begin{figure}[t]
\includegraphics[width=85mm]{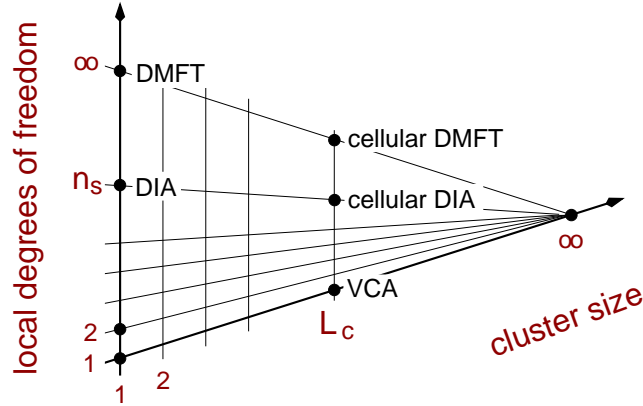}
\centering
\caption{
Dynamical impurity and cluster approximations generated by different reference systems within SFT.
$L_c$ is the number of correlated sites in the reference cluster.
$n_{\rm s}=1+L_b$ is the number of local degrees of freedom where $L_b$ denotes the number of additional bath sites attached to each of the $L_c$ correlated sites.
The variational cluster approximation (VCA) is obtained for finite clusters with $L_c>1$ but without bath sites $n_{\rm s}=1$. 
This generalizes the Hubbard-I-type approximation obtained for $L_c=1$.
The dynamical impurity approximation (DIA) is obtained for $n_{\rm s}>1$ but for a single correlated site $L_c=1$.
A continuum of bath sites, $n_{\rm s}=\infty$ generates DMFT ($L_c=1$) and cellular DMFT ($L_c>1$).
$L_c = \infty$, irrespective of the number of local degrees of freedom, corresponds to the exact solution.
}
\label{fig:approx}
\end{figure}

As a mean-field approach, the DIA does not include the feedback of non-local two-particle correlations on the one-particle spectrum and on the thermodynamics. 
The DIA self-energy is local and takes into account temporal correlations only.
A straightforward idea to include short-range spatial correlations in addition, is to proceed to a reference system with $L_c>1$, i.e.\ system of disconnected finite clusters with $L_c$ sites each. 
The resulting approximation can be termed a cluster mean-field theory since despite the inclusion of short-range correlations, the approximation is still mean-field-like on length scales exceeding the cluster size. 

\index{cellular DMFT}
For an infinite number of bath degrees of freedom $L_b \to \infty$ attached to each of the $L_c >1$ correlated sites the cellular DMFT \cite{KSPB01,LK00} is recovered, see \reffig{approx}.
Considering a single-band Hubbard model again, this can be seen from the corresponding Euler equation: 
\be
   T \sum_{n} \sum'_{ij\sigma}
   \left( 
   \frac{1}{{\bf G}_{0}^{-1}(i\omega_n) - {\bf \Sigma}(i\omega_n)} - {\bf G}'(i\omega_n) 
   \right)_{ij\sigma} 
   \frac{\partial \Sigma_{ji\sigma}(i\omega_n)}
   {\partial {{\bf t}'}}
   = 0  \: ,
\labeq{euler2}
\ee
where the prime at the sum over the sites indicates that $i$ and $j$ must belong to the same cluster of the reference system.
Namely, $\Sigma_{ij\sigma}(i\omega_n) = 0$ and also the ``projector'' ${\partial \Sigma_{ji\sigma}(i\omega_n)}/{\partial {{\bf t}'}} = 0$ if $i$ and $j$ belong to different clusters.
This stationarity condition can be fulfilled if 
\be
   \left( 
   \frac{1}{{\bf G}_{0}^{-1}(i\omega_n) - {\bf \Sigma}(i\omega_n)} - {\bf G}'(i\omega_n) 
   \right)_{ij\sigma} 
   = 0  \: .
\labeq{cdmft}
\ee
Note that ${\bf G}'(i\omega_n)$ is a matrix which is labeled as 
${G}'_{ij,kl}(i\omega_n)$ where $i,j=1,...,L_c$ refer to the correlated sites in the cluster while $k,l = 1,...,L_bL_c$ to the bath sites. 
${G}'_{ij}(i\omega_n)$ are the elements of the cluster Green's function on the correlated sites. 
The condition \refeq{cdmft} is just the self-consistency condition of the C-DMFT.

As is illustrated in \reffig{approx}, the exact solution can be obtained with increasing cluster size $L_c\to\infty$ either from a sequence of reference systems with a continuous bath $n_{\rm s}=1+L_b=\infty$, corresponding to C-DMFT, or from a sequence with $n_{\rm s}=1$, corresponding to VCA, or with a finite, small number of bath sites (``cellular DIA''). 
Systematic studies of the one-dimensional Hubbard model \cite{PAD03,BHP08} have shown that the energy gain which is obtained by attaching a bath site is lower than the gain obtained by increasing the cluster.
This suggests that the convergence to the exact solution could be faster on the ``VCA axis'' in \reffig{approx}. 
For a definite answer, however, more systematic studies, also in higher dimensions, are needed.

In any case bath sites help to get a smooth dependence of physical quantities when varying the electron density or the (physical) chemical potential.
The reason is that bath sites also serve as ``charge reservoirs'', i.e.\ during a $\mu$ scan the ground state of the reference cluster may stay in one and the same sector characterized by the conserved total particle number in the cluster while the particle number on the correlated sites and the approximate particle number in the original lattice model evolve continuously \cite{BHP08,BP10}.
This is achieved by a $\mu$-dependent charge transfer between correlated and bath sites.
In addition, (at least) a single bath site per correlated site in a finite reference cluster is also advantageous to include the interplay between local (Kondo-type) and non-local (antiferromagnetic) singlet formation. 
This has been recognized to be important in studies of the Mott transition \cite{BKS+09} in the two-dimensional and of ferromagnetic order in one-dimensional systems \cite{BP10}.
For studies of spontaneous U(1) symmetry breaking, e.g.\ $d$-wave superconductivity in the two-dimensional Hubbard model \cite{SLMT05,AA05,AAPH06a,AAPH06b}, doping dependencies can be investigated without bath sites due to mixing of cluster states with different particle numbers.

\subsection{Translation symmetry}

For any cluster approximation formulated in real space there is an apparent problem: 
Due to the construction of the reference system as a set of decoupled clusters, the trial self-energies do not preserve the translational symmetries of the original lattice. 
Trivially, this also holds if periodic boundary conditions are imposed for the individual cluster.
Transformation of the original problem to reciprocal space does not solve the problem either since this also means to transform a local Hubbard-type interaction into a non-local interaction part which basically couples all $\ff k$ points.

\index{periodized DMFT}
There are different ideas to overcome this problem.
We introduce a ``periodizing'' universal functional 
\be
  \widehat{T}[\ff \Sigma]_{ij} = \frac{1}{L} \sum_{i'j'} \delta_{i-j,i'-j'} \Sigma_{i'j'}
\ee
which maps any trial self-energy onto translationally invariant one.
In reciprocal space this corresponds to the substitution $\Sigma_{\ff k,\ff k'} \to \widehat{T}[\ff \Sigma]_{\ff k,\ff k'} = \delta_{\ff k,\ff k'} \Sigma_{\ff k}$.
Using this, we replace the self-energy functional of \refeq{sfp} by
\be
  \widehat{\Omega}^{(1)}_{\ff t', \ff U}[\ff \Sigma] = 
  \Tr \ln \frac{1}{\ff G_{\ff t',0}^{-1} - {\widehat \ff T}[\ff \Sigma]}
  + \widehat{F}_{\ff U}[\ff \Sigma] \: ,
\labeq{sfpnew}
\ee
as suggested in Ref.\ \cite{KD05}.
This new functional is different from the original one. 
However, as the physical self-energy is supposed to be translational invariant, it is a stationary point of both, the original and the modified functional.
This means that the modified functional can likewise be used as a starting point to construct approximations.
It turns out (see Ref.\ \cite{PB07} for an analogous discussion in case of disorder) that for a reference system with $L_c>1$ and $n_{\rm s}=\infty$, the corresponding Euler equation reduces to the self-consistency equation of the so-called periodized cellular DMFT (PC-DMFT) \cite{BPK04}.
The same modified functional can also be used to construct a periodized VCA, for example.

\index{dynamical cluster approximation}
While the main idea to recover the PC-DMFT is to modify the form of the self-energy functional, the dynamical cluster approximation (DCA) \cite{HTZ+98,MJPK00,HMJK00} is obtained with the original functional but a modified hopping term in the Hamiltonian of the original system. 
We replace $\ff t \to \widetilde{\ff t}$ and consider the functional $\Omega_{\widetilde{\ff t}, \ff U}[\ff \Sigma]$.
To ensure that the resulting approximations systematically approach the exact solution for cluster size $L_c \to \infty$, the replacement $\ff t \to \widetilde{\ff t}$ must be controlled by $L_c$, i.e.\ it must be exact up to irrelevant boundary terms in the infinite-cluster limit.
This is the case for 
\be
  \widetilde{\ff t} = (\ff V \ff W) \ff U^\dagger \ff t \ff U (\ff V\ff W)^\dagger \: ,
\ee
where $\ff U$, $\ff V$, and $\ff W$ are unitary transformations of the one-particle basis. 
$\ff U$ is the Fourier transformation with respect to the original lattice consisting of $L$ sites ($L\times L$ unitary matrix).
$\ff W$ is the Fourier transformation on the cluster ($L_c \times L_c$), and 
$\ff V$ the Fourier transformation with respect to the superlattice consisting of $L/L_c$ supersites given by the clusters ($L/L_c \times L/L_c$).
The important point is that for any finite $L_c$ the combined transformation $\ff V \ff W = \ff W \ff V \ne \ff U$, while this becomes irrelevant in the limit $L_c\to \infty$.
The detailed calculation (see Ref.\ \cite{PB07} for the analogous disorder case) shows that the DCA is recovered for a reference system with with $L_c>1$ and $n_{\rm s}=\infty$, if periodic boundary conditions are imposed for the cluster.
The same modified construction can also be used to a get simplified DCA-type approximation without bath sites, for example.
This ``simplified DCA'' is related to the periodized VCA in the same way as the DCA is related to the PC-DMFT.
The simplified DCA would represent a variational generalization of a non-self-consistent approximation (``periodic CPT'') introduced recently \cite{MT06}.

\section{Systematics of approximations}

Since the SFT unifies different dynamical approximations within a single formal framework, the question arises how to judge on the relative quality of two different approximations resulting from two different reference systems.
This, however, is not straightforward for several reasons.
First, it is important to note that a stationary point of the self-energy functional is not necessarily a minimum but rather a saddle point in general (see Ref.\ \cite{Pot03a} for an example).
The self-energy functional is not convex.
Actually, despite several recent efforts \cite{Kot99,CK01,NST08}, there is no functional relationship between a thermodynamical potential and time-dependent correlation functions, Green's functions, self-energies, etc.\ which is known to be convex.
\index{convex functionals}

Furthermore, there is no {\em a priori} reason why, for a given reference system, the SFT grand potential at a stationary point should be lower than the SFT grand potential at another one that results from a simpler reference system, e.g.\ a smaller cluster.
This implies that the SFT does not provide upper bounds to the physical grand potential. 
There is e.g.\ no proof (but also no counterexample) that the DMFT ground-state energy at zero temperature must be higher than the exact one. 
On the other hand, in practical calculations the upper-bound property is usually found to be respected, as can be seen for the VCA in Fig.\ \ref{fig:u48}, for example.
Nevertheless, the non-convexity must be seen as a disadvantage as compared to methods based on wave functions which via the Ritz variational principle are able to provide strict upper bounds.

To discuss how to compare two approximations within SFT, we first have to distinguish between ``trivial'' and ``non-trivial'' stationary points for a given reference system. 
A stationary point is referred to as ``trivial'' if the one-particle parameters are such that the reference system decouples into smaller subsystems.
If, at a stationary point, all degrees of freedom (sites) are still coupled to each other, the stationary point is called ``non-trivial''.
It is possible to prove the following theorem \cite{Pot06a}:
Consider a reference system with a set of variational parameters $\ff t' = \ff t'' + \ff V$ where $\ff V$ couples two separate subsystems.
For example, $\ff V$ could be the inter-cluster hopping between two subclusters with completely decouples the degrees of freedom for $\ff V=0$ and all $\ff t''$.
Then,
\be
  \Omega_{\ff t,\ff U}[\ff \Sigma_{\ff t'' + \ff V}] = \Omega_{\ff t,\ff U}[\ff \Sigma_{\ff t''+0}] + {\cal O}(\ff V^2) \; ,
\ee
provided that the functional is stationary at $\ff \Sigma_{\ff t'',\ff U}$ {\em when varying $\ff t''$ only}  (this restriction makes the theorem non-trivial).
This means that going from a more simple reference system to a more complicated one with more degrees of freedom coupled, should generate a new non-trivial stationary point with $\ff V\ne 0$ while the ``old'' stationary point with $\ff V=0$ being still a stationary point with respect to the ``new'' reference system. 
Coupling more and more degrees of freedom introduces more and more stationary points, and none of the ``old'' ones is ``lost''.

Consider a given reference system with a non-trivial stationary point and a number of trivial stationary points.
An intuitive strategy to decide between two stationary points would be to always take the one with the lower grand potential $\Omega_{\ff t,\ff U}[\ff \Sigma_{\ff t',\ff U}]$.
A sequence of reference systems (e.g. $H'_{\rm A}$, $H'_{\rm B}$, $H'_{\rm C}$, ...) in which more and more degrees of freedom are coupled and which eventually recovers the original system $H$ itself, shall be called a ``systematic'' sequence of reference systems.
For such a systematic sequence, the suggested strategy trivially produces a series of stationary points with monotonously decreasing grand potential.
Unfortunately, however, the strategy is useless because it cannot ensure that a systematic sequence of reference systems generates a systematic sequence of approximations as well, i.e.\ one cannot ensure that the respective {\em lowest} grand potential in a systematic sequence of reference systems converges to the exact grand potential.
Namely, the stationary point with the lowest SFT grand potential could be a trivial stationary point (like one associated with a very simple reference system only as $H'_{\rm A}$ or $H'_{\rm B}$ in \reffig{ref}, for example).
Such an approximation must be considered as poor since the exact conditional equation for the self-energy is projected onto a very low-dimensional space only.

Therefore, one has to construct a different strategy which necessarily approaches the exact solution when following up a systematic sequence of reference systems.
Clearly, this can only be achieved if the following rule is obeyed:
{\em A non-trivial stationary point is always preferred as compared to a trivial one (R0).}
A non-trivial stationary point at a certain level of approximation, i.e.\ for a given reference system becomes a trivial stationary point on the next level, i.e.\ in the context of a ``new'' reference system that couples at least two different units of the ``old'' reference system.
Hence, by construction, the rule R0 implies that the exact result is approached for a systematic series of reference systems.

\begin{figure}[t]
\centerline{\includegraphics[width=0.95\textwidth]{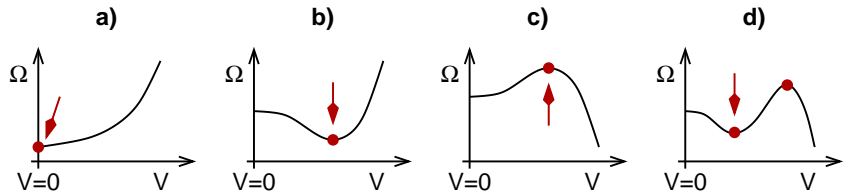}}
\caption{
Possible trends of the SFT grand potential $\Omega$ as a function of a variational parameter $V$ coupling two subsystems of a reference system. 
$V=0$ corresponds to the decoupled case and must always represent a ``trivial'' stationary point.
Circles show the stationary points to be considered.
The point $V=0$ has to be disregarded in all cases except for a).
The arrow marks the respective optimum stationary point according to the rules discussed in the text.
}
\label{fig:om}
\end{figure}

Following the rule (R0), however, may lead to inconsistent thermodynamic interpretations in case of a trivial stationary point with a lower grand potential as a non-trivial one.
To avoid this, R0 has to be replaced by:
{\em Trivial stationary points must be disregarded completely unless there is no non-trivial one (R1).}
This automatically ensures that there is at least one stationary point for any reference system, i.e.\ at any approximation level.

To maintain a thermodynamically consistent picture in case that there are more than a single {\em non-trivial} stationary points, we finally postulate:
{\em Among two non-trivial stationary points for the same reference system, the one with lower grand potential has to be preferred (R2).}

The rules are illustrated by Fig.\ \ref{fig:om}.
Note that the grand potential away from a stationary point does not have a direct physical interpretation.
Hence, there is no reason to interprete the solution corresponding to the maximum in Fig.\ \ref{fig:om}, c) as ``locally unstable''.
The results of Ref.\ \cite{Pot03b} (see Figs.\ 2 and 4 therein) nicely demonstrate that with the suggested strategy (R1, R2) one can consistently describe continuous as well as discontinuous phase transitions.

The rules R1 and R2 are unambiguously prescribed by the general demands for systematic improvement and for thermodynamic consistency.
There is no acceptable alternative to this strategy.
The strategy reduces to the standard strategy (always taking the solution with lowest grand potential) in case of the Ritz variational principle because here a non-trivial stationary point does always have a lower grand potential as compared to a trivial one.

There are also some consequences of the strategy which might be considered as disadvantageous but must be tolerated:
(i) For a sequence of stationary points that are determined by R1 and R2 from a systematic sequence of reference systems, the convergence to the corresponding SFT grand potentials is not necessarily monotonous.
(ii) Given two different approximations specified by two different reference systems, it is not possible to decide which one should be regarded as superior unless both reference systems belong to the same systematic sequence of reference 
systems.
In Fig.\ \ref{fig:ref}, one has $H'_{\rm A} < H'_{\rm B} < H'_{\rm C} < H'_{\rm D}$ where ``$<$'' stands for ``is inferior compared to''.
Furthermore, $H'_{\rm E} < H'_{\rm F}$ and $H'_{\rm A} < H'_{\rm E}$ but there is no relation between $H'_{\rm B}$ and $H'_{\rm E}$, for example.

\section{Summary}

The above discussion has shown that self-energy-functional theory provides a general framework which allows to construct different dynamical approximations for lattice models of strongly correlated electrons.
These approximations derive from a fundamental variational principle, formulated in terms of the grand potential expressed as a functional of the self-energy, by restricting the domain of the functional. 
This leads to non-perturbative and thermodynamically consistent approximations. 
The SFT unifies different known approximations in a single theoretical frame and provides new dynamical impurity (DIA) and variational cluster approximations (VCA).

The essential step in the numerical evaluation consists in the calculation of the Green's function or the self-energy of a reference system with the same interaction part as the original model but with spatially decoupled degrees of freedom.
Details of the numerical procedure can be found in Refs.\ \cite{Pot03a,Pot03b,Sen08,BHP08,BP10}, for example.
Typically, exact diagonalization or the (band) Lanczos approach \cite{LG93,Fre00} but also quantum Monte-Carlo techniques may be used \cite{LHR+09} as a reference-system solver. 
Since bath sites can be integrated out within Green-function based QMC schemes, QMC as an impurity/cluster solver is the method of choice for finite-temperature DMFT or cluster DMFT approaches, i.e.\ for reference systems with a continuum of bath sites.
At zero temperature, and using reference systems without bath sites or a few bath degrees of freedom only, the SFT provides computationally fast techniques which complement the (cluster) DMFT methods.

Besides applications to Hubbard-type model systems, the VCA has recently been employed to study the correlated electronic structure of real materials, such as NiO \cite{Ede07}, CoO, MnO \cite{Ede08}, LaCoO$_3$ \cite{Ede09},
TiOCl \cite{ASV+09}, CrO$_2$ \cite{CAA+07}, TiN \cite{ACA09}, and NiMnSb \cite{ACA+10}.
Furthermore, the theory has been extended to study Bose systems \cite{KD05,KAvdL10a} and the Jaynes-Cummings lattice \cite{AHTL08,KAvdL10b}, electron-phonon systems \cite{KMOH04}, systems with non-local interactions \cite{Ton05}, systems with quenched disorder \cite{PB07} and more.

\subsection*{Acknowledgements}

The author would like to thank
M. Aichhorn,
F.F. Assaad,
E. Arrigoni,
M. Balzer,
R. Bulla,
C. Dahnken,
R. Eder,
W. Hanke,
A. Hewson,
M. Jarrell,
M. Kollar,
G. Kotliar,
A.I. Lichtenstein,
A.J. Millis,
W. Nolting,
D. Senechal,
A.-M.S. Tremblay, 
D. Vollhardt
for cooperations and many helpful discussions.
Support by the Deutsche Forschungsgemeinschaft within the SFB 668 (project A14) and FOR 1346 (project P1) is gratefully acknowledged.


\printindex

\end{document}